\documentclass[preprint,12pt]{elsarticle}

\usepackage[T1]{fontenc}
\usepackage[utf8]{inputenc}
\usepackage{amsmath,amssymb,mathtools,amsthm}
\usepackage{bm}
\usepackage{graphicx}
\usepackage{booktabs,array,tabularx}
\usepackage{enumitem}
\usepackage{needspace}
\usepackage{microtype}
\usepackage{xurl}
\usepackage[hidelinks]{hyperref}
\usepackage{orcidlink}
\hypersetup{pdfauthor={Ruixin Zhou and Boliang Yu}}

\date{}
\makeatletter
\let\ps@pprintTitle\ps@plain
\makeatother
\biboptions{numbers,sort&compress}

\newtheorem{theorem}{Theorem}[section]
\newtheorem{lemma}[theorem]{Lemma}
\newtheorem{proposition}[theorem]{Proposition}
\newtheorem{corollary}[theorem]{Corollary}
\graphicspath{{./}}

\begin{document}

\begin{frontmatter}

\title{Explicit higher local charges in the spin-1 Takhtajan--Babujian chain:
Density-gauge dependence of fixed-boundary MPO complexity}

\author[sjtu]{Ruixin Zhou\,\orcidlink{0009-0004-9303-6674}}
\author[sjtu]{Boliang Yu\,\orcidlink{0009-0006-4664-5068}\corref{cor1}}
\ead{yuboliang@sjtu.edu.cn}
\cortext[cor1]{Corresponding author.}
\address[sjtu]{School of Physics and Astronomy, Shanghai Jiao Tong University, Shanghai 200240, China}

\begin{abstract}
We derive explicit conserved charges through \(Q_5\) for the isotropic spin-1 Takhtajan--Babujian chain and determine how density gauge affects exact matrix-product-operator (MPO) complexity. Among all Hermitian four-site densities obtained from a fixed \(Q_4\) representative by three-site divergences, the minimum full-series MPO dimension is \(36\). Requiring the local density to be reflection even raises this minimum to \(44\). These results concern uniform unblocked one-site tensors with length-independent fixed boundaries and prescribed zero coefficients below length four. The optimization includes non-scalar densities, with explicit SU(2) scalars attaining both bounds. A graded Hankel argument and gauge-invariant coefficient identities establish optimality. An affine family illustrates the distinction between representative and gauge-class minima: its dimensions are \(38\) at two reflection-related points and \(46\) elsewhere. A constructive six-site divergence identity connects the compact five-site charge to the conditional boost recursion for every \(L\ge6\), with \(L=5\) handled separately. Transfer normalization and short-ring completion establish the commuting \(Q_3\)--\(Q_5\) hierarchy. The width-four SU(2)-scalar conserved quotient, modulo divergences and lower charges, is one-dimensional. The same Hankel argument gives exact full-series dimension \(86\) for each of three specified five-site representatives, with coefficients below length five set to zero. Thus a local symmetry constraint has an exact representation cost even though the periodic conserved charge is unchanged.
\end{abstract}


\end{frontmatter}

\section{Introduction}\label{sec:introduction}

Local conserved quantities are a structural signature of quantum integrability and practical inputs for transport, generalized ensembles, and operator-based simulation. In lattice models with a regular Yang--Baxter structure, commuting transfer matrices generate a formal hierarchy, while boost or master-symmetry constructions organize its local levels \cite{R01,R02,R03,R04,R05,R06,R07,R08,R09}. Formal existence, however, does not by itself give compact higher densities that can be inspected, normalized across conventions, or implemented as tensor networks.

The isotropic spin-1 Takhtajan--Babujian chain has well-established Bethe-ansatz and Yang--Baxter foundations \cite{R01,R02,R03,R04}, and classifications identify it as an exceptional integrable point of the bilinear--biquadratic family \cite{R05,R06,R12,R13}. Under the locality assumptions of Hokkyo's theorem, conservation of the third canonical charge generates a commuting boost hierarchy \cite{R42}. Yet explicit higher finite-support densities, their transfer normalization and small-ring completion, and the representation cost of a chosen local density have remained unresolved for this model.

Throughout, lower-case \(q\) denotes a local density and upper-case \(Q\) its periodic translation sum. Adding a local divergence changes \(q\) but leaves \(Q\) invariant; adding a lower conserved charge instead changes the hierarchy basis within the commuting span. The first freedom disappears under periodic summation but survives in a truncated bulk sum through its boundary data. This leads to the central question of the paper: \emph{does local-density gauge freedom leave the exact tensor-network realization cost of a periodic conserved charge invariant?}

The answer is no, and the distinction can be quantified over a complete fixed-width gauge class. Among all Hermitian four-site densities obtained from the fixed \(Q_4\) density by a three-site divergence, the minimum full-series dimension is \(36\). If the local density must be reflection even, the minimum is \(44\). The optimization includes non-scalar densities, with explicit SU(2) scalars attaining both minima. These results apply to uniform unblocked one-site MPOs with length-independent fixed boundaries and prescribed zero coefficients below length four. The periodic charge remains reflection even throughout the class: the extra eight auxiliary states enforce symmetry of its local representative.

The gauge-class result is Theorem~\ref{thm:gauge-optima}. Its explicit partial-trace constructions attain the bounds, while gauge-invariant coefficient identities exclude a smaller dimension throughout the stated Hermitian classes. It determines the least representation cost of a local reflection constraint at fixed width and periodic charge.

Three further results establish the charges and the tools used in this optimization. \emph{Explicit hierarchy:} we give local representatives through \(Q_5\), fix their centered logarithmic-transfer normalization, and identify the compact five-site density with the six-site conditional boost output by a constructed local divergence (Lemma~\ref{lem:q5-recursion-identification}). The identity gives periodic equivalence for every \(L\ge6\); independent small-ring checks complete the compact charge at \(L=5\). The construction supplies both an explicit density and its local connection to the boost hierarchy.

\emph{Low-range classification:} among translation-invariant SU(2)-scalar densities of width at most four satisfying the aligned local continuity condition, the conserved quotient modulo divergences and lower-charge directions is one-dimensional, reflection even, and generated by the displayed \(Q_4\) density (Theorem~\ref{thm:width4-scalar-quotient}). This identifies the scalar class within which representative choice remains free.

\emph{Exact realization cost:} a graded Hankel lemma converts local tensor-train (TT) ranks into the exact dimension of a translated truncated-bulk full series with prescribed zeros below the density width (Lemma~\ref{lem:graded-full-series}). It supplies the matching lower bound for the insertion construction and reduces the four-site gauge optimization to local ranks. Applied at fifth order, it gives dimension \(86\) for each of the three specified five-site representatives (Corollary~\ref{cor:q5-full-series}). This last statement concerns each fixed density; the fourth-order theorem additionally optimizes over its complete stated gauge class.

Two convenient reference densities have dimensions \(38\) and \(46\), neither optimal in the larger gauge class. Along the affine family joining them, the minimum is \(38\) at two reflection-related parameter values and \(46\) everywhere else. The two reference densities and the two optimizing densities also have matching individual tail-only minima, when shorter-word coefficients are left free. Gauge optimization of the tail-only and periodic trace-MPO problems remains open.

Explicit low-order and all-order formulas are known for several spin-\(\tfrac12\) chains. Grabowski and Mathieu displayed the XXX hierarchy through \(Q_4\) and \(Q_5\), including lower-charge admixtures, and later work gave closed or recursive constructions for the XYZ and XXZ chains \cite{R07,R08,R10,R11}. Those expressions do not transfer directly to the present model. The spin-\(\tfrac12\) XXX construction uses the Pauli anticommutator, while the spin-1 rewriting of the SU(3) fundamental chain has local polynomial \(A+A^2\), rather than the Takhtajan--Babujian polynomial \(A-A^2\) \cite{R08}. For the spin-1 Zamolodchikov--Fateev chain, transfer matrices generate the standard hierarchy and additional quasilocal families exist away from the isotropic point \cite{R15}. A recent study of the isotropic Takhtajan--Babujian chain displayed \(Q_3\), related it to a dressed chiral energy current, and left higher charges as a future direction \cite{R14}.

A finite-support density admits a finite-state MPO construction, but this is only an upper bound until a matching lower bound is established \cite{R23,R24,R25,R26,R27,R28,R29,R30,R31}. Tensor-train and operator-Schmidt methods characterize local tensors and fixed-length operators \cite{R28,R29,R30,R31}; Hankel-rank and linear-realization methods instead characterize an all-length coefficient series \cite{R32,R33,R34,R35,R36,R37,R38,R39,R40}. These frameworks also separate uniform fixed-boundary, fixed-length, and periodic trace-MPO optimization. Existing explicit-charge MPOs concern spin-\(\tfrac12\) XYZ/XXZ/XXX or fundamental SU(\(N\)) chains, rather than the spin-1 Takhtajan--Babujian Hamiltonian \cite{R31,R32,R33,R34}.

Here the term \emph{uniform fixed-boundary} means that one site tensor and one pair of boundary vectors must generate the entire operator family for every length. This reusable setting constructs an MPO template once and applies it across system sizes. Requiring one common tensor family across all lengths is more restrictive than optimizing a separate, possibly site-dependent MPO at each fixed length. For the two reference densities, uniform optimization distinguishes dimensions \(38\) and \(46\), whereas the peak fixed-length bond dimension is \(26\) for both once \(L\ge5\). The gauge-class minima \(36\) and \(44\) concern different representatives in the same uniform realization class.

\begin{figure}[!t]
\centering
\includegraphics[width=\textwidth]{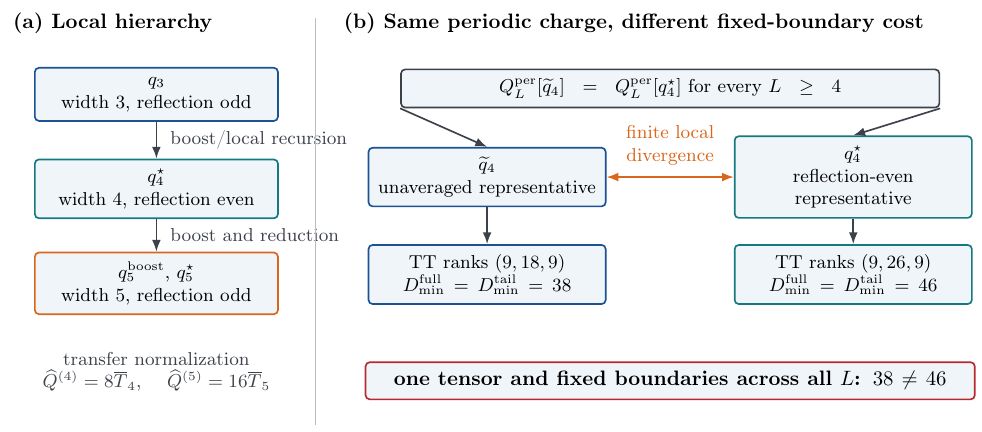}
\caption{Local-density gauge and boundary-sensitive realization cost for two reference densities. (a) The hierarchy through \(Q_5\) fixes the integrable setting and transfer normalization. (b) The densities \(\widetilde q_4\) and \(q_4^\star\) differ by a local divergence and generate the same periodic \(Q_4\). Their full-series and individual tail-only minima are \(38\) and \(46\), respectively. These are representative-specific values; optimization over all Hermitian four-site divergence gauges gives \(36\), or \(44\) with explicit local reflection symmetry, as shown in Sec.~\ref{sec:gauge-optima}.}
\label{fig:proof-routes}
\end{figure}

\begin{figure}[tbp]
\centering
\includegraphics[width=\textwidth]{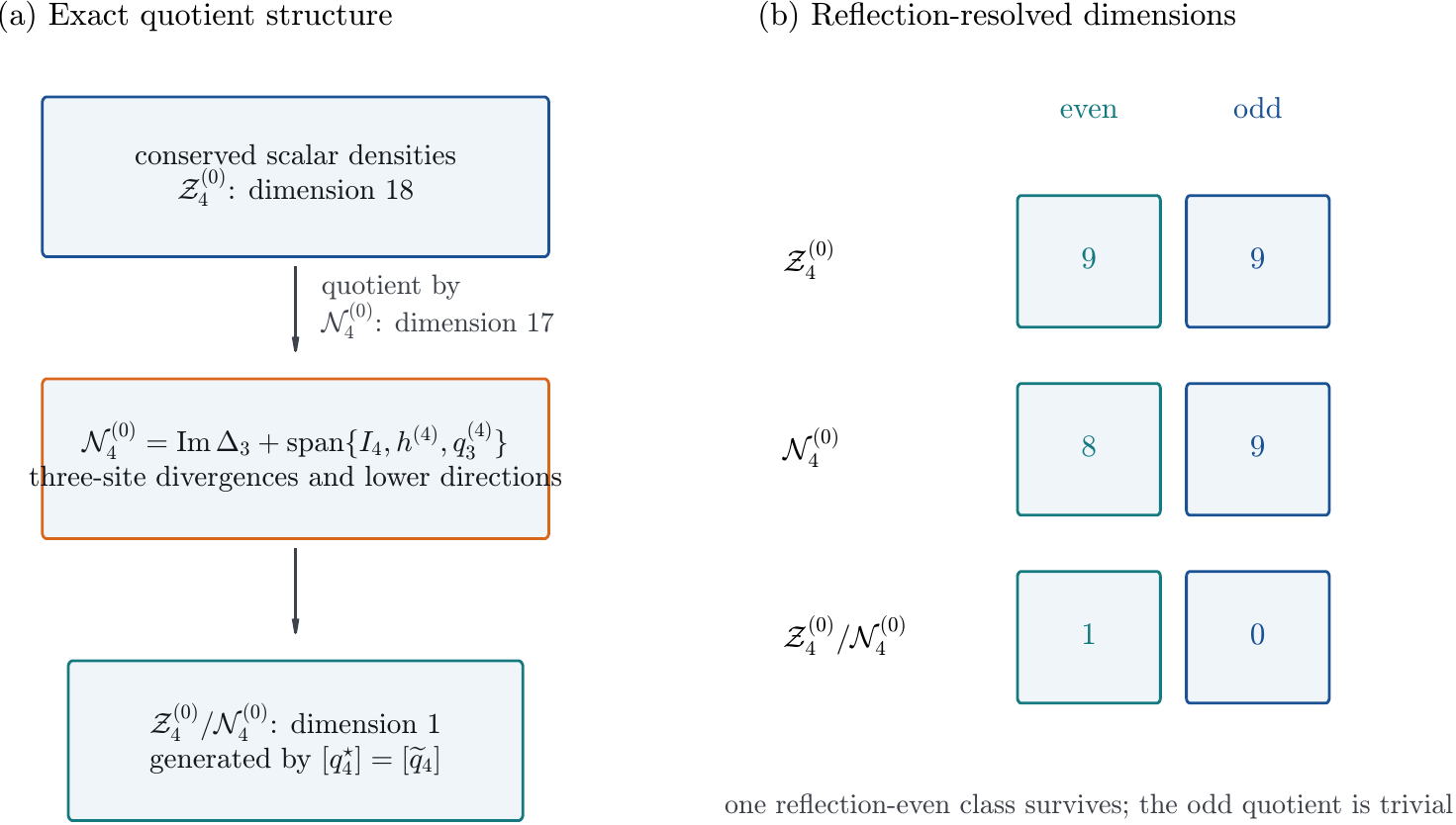}
\caption{Exact width-four scalar conserved-density quotient. (a) The conserved-density space \(\mathcal Z_4^{(0)}\) has dimension \(18\). Quotienting by \(\mathcal N_4^{(0)}=\operatorname{Im}\Delta_3+\operatorname{span}\{I_4,h^{(4)},q_3^{(4)}\}\), of dimension \(17\), leaves the single class generated by \([q_4^\star]=[\widetilde q_4]\). (b) Reflection resolves the exact dimensions as \((9,9)\), \((8,9)\), and \((1,0)\) in the even and odd sectors, respectively. Thus the odd quotient is trivial and the unique surviving class is reflection even.}
\label{fig:width4-scalar-quotient}
\end{figure}

The sources surveyed here do not report four- or five-site Takhtajan--Babujian densities matching the representatives below, the stated width-four classification, or the exact uniform fixed-boundary minima \cite{R07,R08,R14,R15,R31,R32,R33,R34}. This novelty statement is limited to the cited literature. Figure~\ref{fig:proof-routes} displays the periodic-equivalence/fixed-boundary-complexity separation together with the explicit hierarchy that supplies its integrable setting. Figure~\ref{fig:width4-scalar-quotient} gives the scalar quotient behind the low-range uniqueness statement.

Some proofs combine analytic identities with exact rational or integer rank calculations. The appendices provide the technical expansions and computational details. Section~\ref{sec:model} fixes the model and transfer convention; Sec.~\ref{sec:local-continuity} develops local continuity and density gauge; Sections~\ref{sec:q4} and \ref{sec:q5} construct and classify the charges; and Sections~\ref{sec:mpo-construction} and \ref{sec:mpo-minimality} establish the MPO upper and lower bounds. The discussion and conclusions then return to the physical meaning and scope of the result.
\section{Model, transfer matrix, and conventions}\label{sec:model}

\subsection{Spin-1 Hamiltonian and local conventions}

We use the one-site spin basis \(\{|+1\rangle,|0\rangle,|-1\rangle\}\), with \(\hbar=1\), and denote the spin-1 generators by \(\boldsymbol S=(S^x,S^y,S^z)\). On two neighboring sites define

\begin{equation}\label{eq:local-h}
X_{j,j+1}=\boldsymbol S_j\cdot\boldsymbol S_{j+1},
\qquad
h_j=X_{j,j+1}-X_{j,j+1}^{2}.
\end{equation}
The periodic Hamiltonian is

\begin{equation}\label{eq:hamiltonian}
H_L=\sum_{j\in\mathbb Z_L}h_j,
\end{equation}
where site labels are understood modulo \(L\). An overall energy scale and additive constant are fixed by this definition.

For an \(r\)-site density \(q\), let \(\tau_j(q)\) denote its translation to sites \(j,\ldots,j+r-1\). We distinguish the periodic translation sum

\begin{equation}\label{eq:periodic-sum}
Q_{L}^{\mathrm{per}}[q]
=\sum_{j\in\mathbb Z_L}\tau_j(q)
\end{equation}
from the truncated bulk sum

\begin{equation}\label{eq:bulk-sum}
\mathcal B_{r,L}[q]
=\sum_{j=1}^{L-r+1}\tau_j(q).
\end{equation}
The latter contains all translates that fit within an open interval, but it is not called an open-boundary conserved charge unless explicit boundary corrections are supplied.

Reflection of an \(r\)-site operator is denoted by \(\mathcal R_r q\mathcal R_r\), where \(\mathcal R_r\) reverses the site order. Throughout the paper the Hilbert--Schmidt inner product is unnormalized:

\[
\langle A,B\rangle_{\mathrm{HS}}
=\operatorname{Tr}(A^\dagger B).
\]

Whenever a normalized trace is used, it will be stated explicitly.

\subsection{\texorpdfstring{Fused \(R\) matrix}{Fused R matrix}}

Let \(P_s\) project two spin-1 sites onto total spin \(s=0,1,2\). The polynomial fused \(R\) matrix used below is

\begin{equation}\label{eq:fused-R}
R(u)
=(u-1)(u-2)P_0
+(u+1)(u-2)P_1
+(u+1)(u+2)P_2.
\end{equation}
It satisfies the Yang--Baxter equation and the regularity condition

\[
R(0)=2P,
\]

where \(P=P_0-P_1+P_2\) is the two-site permutation. Channel-by-channel differentiation gives the precise relation

\begin{equation}\label{eq:R-to-h}
\boxed{
h=2R^{-1}(0)R'(0)-3I
}.
\end{equation}
Thus the logarithmic derivative reproduces the Hamiltonian density up to the fixed additive constant. The explicit spin matrices, projectors, and a polynomial Yang--Baxter verification are collected in \ref{app:spin1}.

\subsection{\texorpdfstring{Transfer hierarchy and the meaning of \(Q_5\)}{Transfer hierarchy and the meaning of Q5}}

For a periodic chain define

\begin{equation}\label{eq:transfer-matrix}
t_L(u)=\operatorname{tr}_a
R_{aL}(u)\cdots R_{a1}(u).
\end{equation}
Commutativity of the transfer matrices at different spectral parameters gives the usual hierarchy of logarithmic derivatives. Our normalization of \(Q_2\) is the Hamiltonian above, and our \(Q_3\) and \(Q_4\) conventions are fixed by the local formulas in Secs.~\ref{sec:local-continuity} and \ref{sec:q4}.

To prevent an indexing ambiguity, define the \(n\)th transfer charge by

\begin{equation}\label{eq:transfer-charge-definition}
T_{n,L}:=
\left.\partial_u^{\,n-1}
\log\!\left[t_L(u)t_L(0)^{-1}\right]\right|_{u=0},
\qquad n\ge2,
\end{equation}

and remove its scalar part by

\begin{equation}\label{eq:transfer-charge-centered}
\overline T_{n,L}:=
T_{n,L}-\frac{\operatorname{Tr}T_{n,L}}{3^L}I_L.
\end{equation}

Thus the fifth transfer charge \(T_{5,L}\) is the fourth spectral derivative, not the literal fifth derivative. Since all coefficients of \(t_L(u)\) commute, the logarithm in Eq.~\eqref{eq:transfer-charge-definition} is an unambiguous formal power series at the regular point. The five-site objects in Sec.~\ref{sec:q5} are first obtained from the bulk boost and independently proved conserved; Proposition~\ref{prop:hierarchy-normalization} then identifies their exact scale and lower-charge content relative to \(\overline T_{5,L}\).

\subsection{Reshetikhin identity}

Expansion of the Yang--Baxter equation, or direct exact multiplication on three spin-1 sites, yields the model-specific form of Reshetikhin's necessary condition \cite{R05}:

\begin{equation}\label{eq:reshetikhin}
\boxed{
[h_j+h_{j+1},[h_j,h_{j+1}]]
=h_j^2-h_{j+1}^2
}.
\end{equation}
This identity is the local input for the first continuity equation and the subsequent boost construction.

\subsection{Bulk boost and the finite-ring cut}

In the formal local algebra on an infinite chain we write

\begin{equation}\label{eq:boost-hierarchy}
B=\sum_j jh_j,
\qquad
Q_{r+1}=-i[B,Q_r].
\end{equation}
On a finite periodic ring the weighted sum \(\sum_j jh_j\) depends on a cut. Its commutator contains boundary terms that cross the chosen cut; we call these terms the \emph{cut defect}. The weighted sum is therefore not itself a translation-invariant periodic generator, and we use the boost only after separating absolute from relative coordinates in the local algebra. Here \emph{support aliasing} means that periodic identification makes nominally distinct sites within a local identity coincide. Periodic conservation of the resulting densities is proved by local continuity identities and, where such aliasing occurs, by separate exact small-ring witnesses.

For orientation, the representatives used below obey
\[
\widetilde q_4\sim_{\mathrm{div}}q_4^\star,
\qquad
q_5^{\mathrm{boost}}\sim_{\mathrm{div}}q_5^{A},
\qquad
q_5^\star=q_5^{A}-\frac{551}{396}q_{3,5}^{A},
\qquad
Q_{5,L}^{\mathrm{tr}}=-\frac49 i\,\overline T_{5,L}.
\]
The symbol \(\sim_{\mathrm{div}}\) denotes equality up to a finite local divergence and hence equality of periodic translation sums. The subtraction defining \(q_5^\star\), by contrast, changes the basis within the commuting hierarchy; \(q_{3,5}^{A}\) is the orbit-aligned five-site embedding of \(q_3\).

\section{Local continuity and conditional bulk construction}\label{sec:local-continuity}

\subsection{The three-site charge}\label{sec:q3-continuity}

Define

\begin{equation}\label{eq:q3-density}
C_j=[h_j,h_{j+1}],
\qquad
q_{3,j}=iC_j.
\end{equation}
This convention differs by a sign from \cite{R14}. That work uses the same formal boost coordinate \(B=\sum_j jh_j\) but defines \(Q_{m+1}^{\mathrm{R14}}=i[B,Q_m^{\mathrm{R14}}]\), so that

\[
Q_3^{\mathrm{R14}}=-i\sum_j[h_j,h_{j+1}]
=-Q_3^{\mathrm{here}}.
\]

Set \(X_j:=X_{j,j+1}=\boldsymbol S_j\cdot\boldsymbol S_{j+1}\), denoted \(A_j\) in Ref.~\cite{R14}, and \(\chi_j=\boldsymbol S_j\cdot(\boldsymbol S_{j+1}\times \boldsymbol S_{j+2})\). In this notation, Eqs. (17)--(19) of that reference map to

\[
q_{3,j}^{\mathrm{here}}
=\chi_j-\{\chi_j,X_j+X_{j+1}\}
+\{X_j,\{X_{j+1},\chi_j\}\}.
\]

For the physical normalization \(H_{\mathrm{phys}}=(J/4)\sum_jh_j\), where \(J\) is the exchange coupling used in Ref.~\cite{R14}, the corresponding energy-current orientation gives

\[
J_E=\left(\frac J4\right)^2Q_3^{\mathrm{here}}.
\]

Thus the sign of a statement identifying \(Q_3\) with energy current is tied to both the boost and continuity-equation orientations.

Since \(C_j\) is real antisymmetric, \(q_{3,j}\) is Hermitian. Introduce

\[
F_j=[h_{j-1},C_j]+h_j^2.
\]

Using the Reshetikhin identity, the Jacobi identity, and locality,

\begin{equation}\label{eq:q3-continuity}
\begin{aligned}
[H,C_j]
&=[h_{j-1},C_j]+[h_j+h_{j+1},C_j]+[h_{j+2},C_j]\\
&=[h_{j-1},C_j]+h_j^2-h_{j+1}^2-[h_j,C_{j+1}]\\
&=F_j-F_{j+1}.
\end{aligned}
\end{equation}
For periodic embeddings with four distinct sites available to this local identity, the translation sum telescopes. The aliased three-site ring is covered separately by exact integer arithmetic, which also verifies that the charge is nonzero:

\begin{equation}\label{eq:q3-conservation}
\boxed{
[H_L,Q_{3,L}^{\mathrm{per}}]=0,
\qquad L\ge3.
}
\end{equation}
Thus \(L\ge4\) is the unaliased local-proof range, while exact integer arithmetic completes the statement at \(L=3\). The computational methods are summarized in \ref{app:crosswalk}.

\subsection{Conditional bulk recursion}

Suppose an \(r\)-site density satisfies

\begin{equation}\label{eq:continuity-template}
[H,q_j^{(r)}]
=J_j^{(r)}-J_{j+1}^{(r)}.
\end{equation}
Only the bonds beginning at \(j-1,\ldots,j+r-1\) overlap \(q_j^{(r)}\). Separating \(j+s\) into its absolute coordinate \(j\) and relative coordinate \(s\) gives

\[
\begin{aligned}
Q_{r+1}
&=-i\sum_j\sum_{s=-1}^{r-1}(j+s)
[h_{j+s},q_j^{(r)}]\\
&=-i\sum_j j[H,q_j^{(r)}]
-i\sum_j\sum_{s=-1}^{r-1}s[h_{j+s},q_j^{(r)}].
\end{aligned}
\]

The absolute-coordinate term is reduced with

\[
\sum_j j(J_j-J_{j+1})=\sum_jJ_j.
\]

A constructive density at the next boost level is therefore

\begin{equation}\label{eq:conditional-boost-density}
\boxed{
q_{j,\mathrm{rec}}^{(r+1)}
=-i\left[
J_j^{(r)}
+\sum_{s=-1}^{r-1}s[h_{j+s},q_j^{(r)}]
\right].
}
\end{equation}
This recursive construction applies whenever a local current is available. Control of support, symbolic complexity, and MPO dimension at arbitrary order remains a separate problem.

\subsection{Density gauge and hierarchy basis}\label{sec:density-gauge}

Two freedoms must be kept separate. A density-gauge transformation has the form

\[
q_j\longmapsto q_j+D_j-D_{j+1}.
\]

Its periodic translation sum vanishes by telescoping, so it changes the local representative without changing the periodic operator on the stated chain lengths.

A hierarchy-basis transformation instead has the form

\[
Q_r\longmapsto cQ_r+\sum_{k<r}c_kQ_k,
\qquad c\ne0.
\]

It changes the conserved operator within the span of the hierarchy. The subtraction of a \(Q_3\) component from the five-site charge in Sec.~\ref{sec:q5} is of this second type, whereas translation-orbit alignment is a density-gauge choice. This distinction is already visible in the spin-\(\tfrac12\) XXX hierarchy. Grabowski and Mathieu \cite{R07} write \(Q_4=2H_4-4H_2\) and \(Q_5=6H_5-18H_3\). Their \(H_n\) basis strips lower-charge contributions from the direct boost representatives; see also the general discussion in \cite{R08}. Our lower-charge subtraction fixes the comparison convention used below.

\section{\texorpdfstring{Four-site representatives and density gauge}{Four-site representatives and density gauge}}\label{sec:q4}

\subsection{Bulk-boost representative}

On four consecutive sites set

\[
a=h_{12},\qquad b=h_{23},\qquad c=h_{34},
\]

and define

\begin{equation}\label{eq:q4-g}
g=a^2+[b,[a,b]]+2[c,[a,b]].
\end{equation}
Applying the bulk boost to \(Q_3\) and using the current \(F_j\) from Sec.~\ref{sec:q3-continuity} gives

\[
Q_4=\sum_j g_j.
\]

The construction is carried out in the local algebra before periodic embedding; it is not obtained by taking the commutator with an uncorrected finite-ring boost.

\begin{proposition}[Formal boost conservation at fourth order]\label{prop:q4-formal}
In the translation-invariant formal local algebra,

\begin{equation}\label{eq:q4-formal-recursion}
Q_3=-i[B,H],\qquad Q_4=-i[B,Q_3].
\end{equation}
\end{proposition}

\begin{proof}
The continuity equation in Sec.~\ref{sec:q3-continuity} gives \([H,Q_3]=0\) after summing the one-site divergence. Since \([H,B]=-iQ_3\), the Jacobi identity yields

\begin{equation}\label{eq:q4-jacobi}
\begin{aligned}
[H,Q_4]
&=-i[H,[B,Q_3]]\\
&=-i\left([[H,B],Q_3]+[B,[H,Q_3]]\right)\\
&=-i[-iQ_3,Q_3]=0.
\end{aligned}
\end{equation}
\end{proof}

This formal extensive identity is obtained after removing the absolute boost coordinate. We apply it to periodic chains through the local density \(g\).

\subsection{Reflection representative and lower-charge hierarchy convention}

The periodic sums of \(g\) and its reflection agree. We therefore first take the reflection-even representative

\[
e=\frac{g+\mathcal R_4g\mathcal R_4}{2}.
\]

On the four-site window let \(d_H=(a+c)/2\). Exact Hilbert--Schmidt orthogonalization of \(e+\alpha d_H+\beta_4 I\) against the local identity and Hamiltonian directions gives

\[
\alpha=11,
\qquad
\beta_4=\frac{28}{3}.
\]

The sign and overall rational scale are fixed by defining

\[
\boxed{
q_4^\star
=-\frac1{36}\left[
\frac{g+\mathcal R_4g\mathcal R_4}{2}
+\frac{11}{2}(a+c)
+\frac{28}{3}I
\right].
}
\]

Equivalently,

\begin{equation}\label{eq:q4-density}
\boxed{
216q_4^\star
=-\left[
3(g+\mathcal R_4g\mathcal R_4)+33(a+c)+56I
\right].
}
\end{equation}
The lower-charge additions in this formula fix a hierarchy-basis convention. To make the density-gauge statement explicit, retain the same lower-charge convention but omit the reflection average:

\begin{equation}\label{eq:q4-tilde}
\boxed{
216\widetilde q_4
=-\left[6g+33(a+c)+56I\right].
}
\end{equation}
Exact orbit integration produces a three-site integer matrix \(K_{\mathrm{rep}}\). In Eq.~\eqref{eq:q4-representative-divergence}, the first term embeds it on sites \((1,2,3)\) of the four-site window and the second on sites \((2,3,4)\):
\begin{equation}\label{eq:q4-representative-divergence}
6^7(\widetilde q_4-q_4^\star)
=K_{\mathrm{rep}}\otimes I-I\otimes K_{\mathrm{rep}}.
\end{equation}
Here \(6^7=6^4\times216\): the first factor clears the rational product-basis denominators and the second is the physical density normalization. Equation~\eqref{eq:q4-representative-divergence} explicitly realizes the difference as a local divergence and therefore proves equality of the periodic sums for every \(L\ge4\). Their truncated-bulk boundary sectors remain different. We henceforth abbreviate the periodic sums as
\[
\widetilde Q_{4,L}^{\mathrm{per}}:=Q_L^{\mathrm{per}}[\widetilde q_4],
\qquad
Q_{4,L}^{\star,\mathrm{per}}:=Q_L^{\mathrm{per}}[q_4^\star].
\]

\subsection{Exact properties and chain-length scope}

Exact matrix algebra gives

\[
(q_4^\star)^\dagger=q_4^\star,
\qquad
\mathcal R_4q_4^\star\mathcal R_4=q_4^\star,
\]

\[
[S^\alpha_{\mathrm{tot}},q_4^\star]=0,
\qquad
\operatorname{Tr}q_4^\star=0,
\qquad
\|q_4^\star\|_{\mathrm{HS}}^2=39.
\]
\ref{app:q4} gives the projector expansion, orthogonalization data, and exact checks behind this representative.

Reflection averaging does not change the periodic translation sum, while the added Hamiltonian and identity directions commute with \(H_L\). For a finite-ring proof independent of the formal boost coordinate, let \(K_4^{[72]}\) be the five-site integer current matrix at scale \(72\). Direct integer matrix multiplication on the unaliased six-site window gives
\begin{equation}\label{eq:q4-continuity}
\begin{aligned}
\sum_{s=-1}^{3}[h_{j+s},216q_{4,j}^\star]
&=3\left(K_4^{[72]}\otimes I-I\otimes K_4^{[72]}\right),\\
J_j^{(4)}&=\frac{1}{72}(K_4^{[72]})_{j-1,\ldots,j+3}.
\end{aligned}
\end{equation}
After division by \(216\), Eq.~\eqref{eq:q4-continuity} is exactly
\([H,q_{4,j}^\star]=J_j^{(4)}-J_{j+1}^{(4)}\).
For every periodic chain with \(L\ge6\), summing this one-site divergence telescopes without support aliasing. Exact integer-matrix calculations cover \(L=4,5\) and also verify nontriviality. Consequently,
\begin{equation}\label{eq:q4-conservation}
\boxed{
[H_L,Q_{4,L}^{\star,\mathrm{per}}]=0,
\qquad L\ge4.
}
\end{equation}
The same statement holds for \(\widetilde q_4\) by Eq.~\eqref{eq:q4-representative-divergence}. Thus Proposition~\ref{prop:q4-formal} identifies the integrable origin of the density, while the explicit current identity and small-ring calculations establish the stated finite-ring range.

This representative is supported on four consecutive sites. The next result proves uniqueness of its conserved scalar class among densities represented within a four-site window. It does not yet assert that every representative of this class must have width four.

\subsection{Classification of scalar conserved densities through width four}\label{sec:q4-classification}

Let

\[
\mathcal V_w^{(0)}
:=\operatorname{End}_{SU(2)}\!\left[(\mathbb C^3)^{\otimes w}\right]
\]

be the complexified space of SU(2)-scalar \(w\)-site operators. Shorter densities are padded by identities into a four-site window. On a common five-site window labeled \(0,\ldots,4\), write

\[
q_{\mathrm L}:=q\otimes I\quad\text{on sites }0,\ldots,3,
\qquad
q_{\mathrm R}:=I\otimes q\quad\text{on sites }1,\ldots,4.
\]

The five nonvanishing relative placements in the commutator of the Hamiltonian with a four-site density are translated so that their union occupies this window; \(h_{rs}\) denotes \(h\) on sites \(r,s\). For \(q,j\in\mathcal V_4^{(0)}\), the resulting maps are

\begin{equation}\label{eq:width4-classification-maps}
\begin{aligned}
\mathcal C_4(q):={}&
[h_{01},q_{\mathrm R}]+[h_{01},q_{\mathrm L}]
+[h_{12},q_{\mathrm L}]\\
&+[h_{23},q_{\mathrm L}]+[h_{34},q_{\mathrm L}],\\
\Delta_4(j):={}&j\otimes I-I\otimes j.
\end{aligned}
\end{equation}

For definiteness, choose right-padded four-site representatives of the lower directions,

\[
I_4:=I_3^{\otimes4},
\qquad
h^{(4)}:=h\otimes I_3^{\otimes2},
\qquad
q_3^{(4)}:=q_3\otimes I_3.
\]

For \(L\ge4\), their periodic sums are \(LI_L\), \(H_L\), and \(Q_{3,L}^{\mathrm{per}}\), respectively. The scalar locally conserved-density space and the lower-direction subspace are

\begin{align}
\mathcal Z_4^{(0)}
&:=\{q\in\mathcal V_4^{(0)}:\mathcal C_4(q)\in\operatorname{Im}\Delta_4\},
\label{eq:scalar-conserved-space}\\
\mathcal N_4^{(0)}
&:=\Delta_3(\mathcal V_3^{(0)})
+\operatorname{span}\{I_4,h^{(4)},q_3^{(4)}\},
\label{eq:scalar-gauge-lower-space}
\end{align}

Changing the padding chosen above changes only the divergence part. The aligned operator \(\mathcal C_4(q)\) is the five-site local density of \([H,Q^{\mathrm{per}}[q]]\). Membership in \(\operatorname{Im}\Delta_4\) is the one-site local continuity condition used in the classification below. Periodic summation of the corresponding divergence makes the commutator vanish on every unaliased embedding; no converse at an aliased finite length is assumed.

\begin{theorem}[Width-four scalar quotient]\label{thm:width4-scalar-quotient}
For the spin-1 Takhtajan--Babujian density \(h\),

\begin{equation}\label{eq:width4-quotient-dimension}
\boxed{
\dim\mathcal Z_4^{(0)}=18,
\qquad
\dim\mathcal N_4^{(0)}=17,
\qquad
\dim\bigl(\mathcal Z_4^{(0)}/\mathcal N_4^{(0)}\bigr)=1.
}
\end{equation}

Under reflection, the conserved and gauge-and-lower-charge subspace dimensions are respectively

\begin{equation}\label{eq:width4-parity-dimensions}
\begin{array}{c|cc}
&\text{even}&\text{odd}\\ \hline
\mathcal Z_4^{(0)}&9&9\\
\mathcal N_4^{(0)}&8&9\\
\mathcal Z_4^{(0)}/\mathcal N_4^{(0)}&1&0.
\end{array}
\end{equation}

The density \(q_4^\star\), equivalently \(\widetilde q_4\), generates the unique surviving reflection-even class.
\end{theorem}

\begin{proof}
Successive spin-1 coupling gives multiplicities \((1,3,2,1)\) for total spins \(0,1,2,3\) on three sites and \((3,6,6,3,1)\) for total spins \(0,1,2,3,4\) on four sites. Schur decomposition therefore gives

\[
\dim\mathcal V_3^{(0)}=1^2+3^2+2^2+1^2=15,
\qquad
\dim\mathcal V_4^{(0)}=3^2+6^2+6^2+3^2+1^2=91.
\]

An exact rational intertwiner basis is obtained by imposing commutation with the integer spin-1 \(sl_2\) raising and Cartan generators; commutation with the lowering generator is then checked independently. In these bases, the joint map

\[
\mathcal M:\mathcal V_4^{(0)}\oplus\mathcal V_4^{(0)}
\longrightarrow\operatorname{End}[(\mathbb C^3)^{\otimes5}],
\qquad
\mathcal M(q,j)=\mathcal C_4(q)-\Delta_4(j),
\]

is represented by a \(59049\times182\) sparse rational matrix. Exact elimination gives \(\dim\ker\mathcal M=19\). The kernel of \(\Delta_4\) on \(\mathcal V_4^{(0)}\) is the one-dimensional identity-current direction, so projection onto the density component gives \(\dim\mathcal Z_4^{(0)}=19-1=18\).

Likewise, \(\Delta_3\) has rank \(15-1=14\). The padded \(I_4,h^{(4)},q_3^{(4)}\) directions are independent modulo this image, giving \(\dim\mathcal N_4^{(0)}=17\), and direct substitution verifies \(\mathcal N_4^{(0)}\subset\mathcal Z_4^{(0)}\). Reflection projection gives the two rows of Eq.~\eqref{eq:width4-parity-dimensions}. Finally, adjoining \(q_4^\star\) increases the gauge-and-lower-charge rank by one without increasing the conserved rank, so its class generates the quotient. Independent modular calculations reproduce the rational ranks.
\end{proof}

The quotient theorem makes the later MPO comparison intrinsic to the first nontrivial scalar hierarchy class. The two four-site representatives compared below are density gauges of this unique surviving width-four class.

\section{Explicit five-site charges and hierarchy normalization}\label{sec:q5}

The five-site construction has three logically separate steps. We first compress the conditional boost output to a short representative by removing a local divergence. We then choose a consistent placement within each translation orbit, and finally subtract the embedded \(Q_3\) component to fix a local Hilbert--Schmidt hierarchy convention. The first step exposes the local formula; the second changes density gauge without changing the periodic charge; the third makes an explicit, convention-dependent lower-charge subtraction. The transfer-normalized hierarchy direction is identified separately in Proposition~\ref{prop:hierarchy-normalization}.

\subsection{Divergence-reduced ten-word boost representative}

On five sites define

\[
a=h_{12},\qquad b=h_{23},\qquad
c=h_{34},\qquad d=h_{45}.
\]

For a noncommutative bond word \(w\), set

\[
\mathcal O(w)
=(w-w^\dagger)
-\mathcal R_5(w-w^\dagger)\mathcal R_5.
\]

The operator \(\mathcal O(w)\) is real antisymmetric and reflection odd, so \(i\mathcal O(w)\) is Hermitian and reflection odd. After reducing the six-site conditional-recursion output by the finite-support divergence displayed below, the resulting five-site representative has the ten-word-orbit form

\begin{equation}\label{eq:q5-word}
\begin{aligned}
\mathcal P_5^{\mathrm{boost}}={}&
15\mathcal O(aabb)
+72\mathcal O(abcd)
-72\mathcal O(acbd)
-144\mathcal O(abdc)\\
&+72\mathcal O(abbc)
-72\mathcal O(abcb)
-72\mathcal O(babc)\\
&+2\mathcal O(aaba)
+2\mathcal O(babb)
+24\mathcal O(ab),
\end{aligned}
\end{equation}
\begin{equation}\label{eq:q5-boost}
\boxed{
q_5^{\mathrm{boost}}=\frac{i}{864}\mathcal P_5^{\mathrm{boost}}
=\frac{i}{72}A_{72},
}
\end{equation}
where \(A_{72}=\mathcal P_5^{\mathrm{boost}}/12\) is a real antisymmetric integer matrix. The term ten-word counts the ten displayed word-orbit terms, irrespective of linear relations among them in the full operator space.

The relation to the conditional recursion can be established on a single six-site window. Put \(G_{216}=216q_4^\star\) on sites \(2,\ldots,5\), and put the current \(K_4^{[72]}\) of Eq.~\eqref{eq:q4-continuity} on sites \(1,\ldots,5\). Substitution into Eq.~\eqref{eq:conditional-boost-density} gives the explicit numerator
\begin{equation}\label{eq:q5-recursion-explicit}
\begin{aligned}
W_6&:=\sum_{s=-1}^{3}s\,[h_{s+2,s+3},I\otimes G_{216}\otimes I],\\
\widehat A_{72}^{(6)}&=-K_4^{[72]}\otimes I-\frac13W_6,
\qquad q_{5,\mathrm{rec}}^{(6)}=\frac{i}{72}\widehat A_{72}^{(6)}.
\end{aligned}
\end{equation}
Every entry of \(W_6\) is divisible by three. The resulting numerator is therefore an integer matrix obtained directly from the conditional recursion.

Here the left endpoint of \(q_4^\star\) is \(j=2\). The overlapping bonds
start at \(j+s=1,\ldots,5\), so their relative weights are
\(-1,0,1,2,3\). The current in physical units is
\(J^{(4)}=K_4^{[72]}/72\). Thus direct substitution first gives
\[
q_{5,\mathrm{rec}}^{(6)}
=-i\left(\frac{K_4^{[72]}\otimes I}{72}+\frac{W_6}{216}\right).
\]
Factoring out \(i/72\) yields Eq.~\eqref{eq:q5-recursion-explicit};
the factor \(1/3\) is the ratio \(72/216\). The current itself can be
reconstructed from \(G_{216}\) by the orbit integration in
\ref{app:q5-recursion-details}, so this formula specifies the recursion
output without taking the compact five-site expression as an input.

\Needspace{10\baselineskip}
\begin{lemma}[Local identification of the compact fifth charge]\label{lem:q5-recursion-identification}
With the above origins and the traceless current convention, there is a five-site real antisymmetric integer matrix \(K_{\mathrm B}\) for which
\begin{equation}\label{eq:q5-boost-identity}
\boxed{
6^6\left(\widehat A_{72}^{(6)}-A_{72}\otimes I\right)
=K_{\mathrm B}\otimes I-I\otimes K_{\mathrm B}.
}
\end{equation}
Consequently, the periodic translation sums of \(q_{5,\mathrm{rec}}^{(6)}\) and \(q_5^{\mathrm{boost}}\) agree for every \(L\ge6\).
\end{lemma}

\begin{proof}
Expand \(6^6(\widehat A_{72}^{(6)}-A_{72}\otimes I)\) in the rational
product basis of \ref{app:q5}. The recursion numerator is formed from
Eq.~\eqref{eq:q5-recursion-explicit}, and \(A_{72}\) is formed independently
from the ten bond words. For each translation orbit, list the coefficients
of its successive placements as \(c_0,\ldots,c_k\). Exact expansion gives
\(\sum_{t=0}^k c_t=0\). For \(k\ge1\), define
\[
j_t=\sum_{s=0}^t c_s,\qquad 0\le t<k.
\]
The divergence of the current with these coefficients reproduces every
placement coefficient:
\[
j_0=c_0,\qquad j_t-j_{t-1}=c_t\ (1\le t<k),\qquad -j_{k-1}=c_k.
\]
A single-placement orbit has zero coefficient. Summing the constructed
currents over all orbits gives
the five-site matrix \(K_{\mathrm B}\), with its identity component set to
zero. Exact matrix reconstruction verifies Eq.~\eqref{eq:q5-boost-identity}
entry by entry. \ref{app:q5-recursion-details} supplies the coefficient
formula and the construction of the input current from \(G_{216}\).

With \(D=iK_{\mathrm B}/(72\,6^6)\), the physical-density difference is
\(D\otimes I-I\otimes D\). On every ring with \(L\ge6\), the six-site
identity has a well-defined embedding and its translated divergence
telescopes. This proves equality of the periodic sums on all these lengths.
\end{proof}
The compact charge is also well defined on \(L=5\), where its conservation
and commutativity follow from the separate small-ring checks below. The
six-site recursion identity is used on \(L\ge6\).

In particular, the all-length step is the local-to-periodic implication
\begin{equation}\label{eq:q5-local-to-periodic}
\begin{aligned}
q_{5,\mathrm{rec}}^{(6)}
 &=q_5^{\mathrm{boost}}\otimes I+D\otimes I-I\otimes D,\\
Q_L^{\mathrm{per}}[q_{5,\mathrm{rec}}^{(6)}]
 &=Q_L^{\mathrm{per}}[q_5^{\mathrm{boost}}],\qquad L\ge6.
\end{aligned}
\end{equation}
The second line follows because the two sums of translated \(D\) differ only by a cyclic relabeling. The local identity thus connects the compact formula to the boost hierarchy for all \(L\ge6\).

A short orbit illustrates the construction in the proof. In the rational
product basis of \ref{app:q5}, take the four-site core
\(\Gamma=B_5\otimes B_2\otimes B_2\otimes B_6\).
Its three placements in
\(6^6(\widehat A_{72}^{(6)}-A_{72}\otimes I)\) have coefficients
\[
(c_0,c_1,c_2)=69984(2,-1,-1).
\]
Their sum is zero, and the two current coefficients are
\((j_0,j_1)=69984(2,1)\). The contribution
\(69984(2\Gamma\otimes I+I\otimes\Gamma)\) to \(K_{\mathrm B}\)
therefore reproduces all three placements under \(K\mapsto K\otimes I-I\otimes K\).
The same finite calculation is performed for every orbit; the dataset
records all placement coefficients, cumulative sums, and the reconstructed
matrix residual.

\subsection{Translation-orbit alignment and lower-charge fixing}

The compact ten-word formula does not fix where a translated product word is placed inside the five-site window. In the real rational product basis specified in \ref{app:q5}, let \(\mathcal L_r(A)\) collect the total coefficient of every finite-support translation orbit and place its active core at the left edge. We define

\begin{equation}\label{eq:orbit-alignment}
\boxed{
\operatorname{Align}_{\mathrm{orb},r}^{(-)}(A)
=\frac12\left[
\mathcal L_r(A)
-\mathcal R_r\mathcal L_r(A)\mathcal R_r
\right].
}
\end{equation}
This convention is globally non-idempotent, but the three outputs used below are exact fixed points. Full basis definitions and exact diagnostics are given in \ref{app:q5}.

\begin{lemma}[Finite-support density-gauge preservation]\label{lem:alignment-preserves-charge}

Let \(A\) be an \(r\)-site reflection-odd operator expanded in the specified product basis. There is a finite-support operator \(D_A^{(-)}\) such that, with \(T\) denoting one-site right translation in the formal local algebra,

\begin{equation}\label{eq:alignment-divergence}
\boxed{
\operatorname{Align}_{\mathrm{orb},r}^{(-)}(A)-A
=D_A^{(-)}-T D_A^{(-)}.
}
\end{equation}
Consequently, for every periodic embedding with \(L\ge r\),

\begin{equation}\label{eq:alignment-charge}
\boxed{
Q_L^{\mathrm{per}}\!\left[
\operatorname{Align}_{\mathrm{orb},r}^{(-)}(A)
\right]
=Q_L^{\mathrm{per}}[A].
}
\end{equation}
\end{lemma}

\begin{proof}
Within each orbit, moving a term \(T^sY_\gamma\) to its left-aligned position changes it by \((1-T)\sum_{t=0}^{s-1}T^tY_\gamma\). Summing these finite coboundaries gives \(\mathcal L_r(A)-A=(1-T)D_A\). Reflection maps this to another finite coboundary because \(\mathcal RTR=T^{-1}\); antisymmetrizing therefore yields Eq.~\eqref{eq:alignment-divergence}. Periodic summation telescopes and proves Eq.~\eqref{eq:alignment-charge}. The explicit current is recorded in \ref{app:q5}.
\end{proof}

The lemma shows that alignment changes only density gauge. To state the separate lower-charge convention, embed the three-site real antisymmetric numerator centrally in the five-site window:

\[
A_3^{(5)}
=I\otimes[h_{12},h_{23}]\otimes I.
\]

The corresponding Hermitian five-site reference is

\begin{equation}\label{eq:q3-five-site}
\boxed{
q_3^{(5)}:=iA_3^{(5)}
=I\otimes q_3\otimes I.
}
\end{equation}
We name the alignment-only representatives
\[
q_5^{A}:=\operatorname{Align}_{\mathrm{orb},5}^{(-)}
(q_5^{\mathrm{boost}}),
\qquad
q_{3,5}^{A}:=\operatorname{Align}_{\mathrm{orb},5}^{(-)}
(q_3^{(5)}).
\]
A separate operation fixes the lower-charge convention: we remove the \(q_{3,5}^{A}\) component using the local Hilbert--Schmidt product in this aligned five-site window. This basis choice changes the periodic hierarchy element and is not density gauge. The projection coefficient in the Hermitian charge normalization is

\[
\boxed{
\beta_5=\frac{551}{396}.
}
\]

We define

\begin{equation}\label{eq:q5-aligned-hs}
\boxed{
\begin{aligned}
q_5^\star
&:=q_5^{A}-\frac{551}{396}q_{3,5}^{A}\\
&=\frac{i}{132}A_5^\star.
\end{aligned}
}
\end{equation}
Here \(A_5^\star\) is a real antisymmetric integer matrix. In the stated spin-product basis, the greatest common divisor of its nonzero entries is one; this is the meaning of primitive used here.

Lemma~\ref{lem:alignment-preserves-charge} separates the two operations in this definition. Orbit alignment changes only the density representative of each reflection-odd input. Subtracting \(551Q_3/396\), by contrast, is a hierarchy-basis choice and changes the conserved operator:

\begin{equation}\label{eq:q5-hierarchy}
Q_{5,L}^{\star,\mathrm{per}}
=Q_{5,L}^{\mathrm{boost},\mathrm{per}}
-\frac{551}{396}Q_{3,L}^{\mathrm{per}},
\qquad L\ge5.
\end{equation}

Thus \(q_5^A\) is divergence-equivalent to \(q_5^{\mathrm{boost}}\), whereas \(q_5^\star\) additionally fixes a convention-dependent hierarchy basis. Changing the alignment before the local projection can change the lower-charge coefficient. The transfer-normalized direction introduced next is the convention used for comparisons that should not depend on this local choice.

\begin{proposition}[Normalization map to the boost and transfer hierarchies]\label{prop:hierarchy-normalization}
In the infinite-chain formal local algebra, let
\[
h^{(0)}=h+\frac43I_9,
\qquad
H^{(0)}:=\sum_{j\in\mathbb Z}h_j^{(0)},
\qquad
B_{H^{(0)}}:=\sum_{j\in\mathbb Z}j\,h_j^{(0)},
\]
and define Hokkyo's hierarchy convention by
\[
\widehat Q^{(2)}=H^{(0)},
\qquad
\widehat Q^{(n+1)}=[B_{H^{(0)}},\widehat Q^{(n)}].
\]
Let \(\widehat Q_{L}^{(n),\mathrm{per}}\) denote the periodic translation sum of the resulting local density, let \(I_L\) be the identity on the \(L\)-site Hilbert space, and set \(H_L^{(0)}=H_L+4LI_L/3\). The formal boost recursion fixes the non-scalar hierarchy direction; on periodic embeddings we fix its remaining identity component by the traceless transfer convention. The normalization relations are then
\begin{align}
Q_{3,L}^{\mathrm{per}}
&=-i\widehat Q_{L}^{(3),\mathrm{per}},
&&L\ge3,\label{eq:hierarchy-map-q3}\\
\widehat Q_{L}^{(4),\mathrm{per}}
&=8\overline T_{4,L}
=-Q_L^{\mathrm{per}}[g]+\frac{16}{3}LI_L,
&&L\ge4,\label{eq:q4-transfer-map}\\
Q_{4,L}^{\star,\mathrm{per}}=\widetilde Q_{4,L}^{\mathrm{per}}
&=\frac1{36}\widehat Q_{L}^{(4),\mathrm{per}}
-\frac{11}{36}H_L^{(0)},
&&L\ge4,\label{eq:hierarchy-map-q4}\\
Q_{5,L}^{\mathrm{boost},\mathrm{per}}
&=-\frac{i}{36}\widehat Q_{L}^{(5),\mathrm{per}}-\frac{11}{36}Q_{3,L}^{\mathrm{per}},
&&L\ge6,\label{eq:hierarchy-map-q5boost}\\
Q_{5,L}^{\star,\mathrm{per}}
&=-\frac{i}{36}\widehat Q_{L}^{(5),\mathrm{per}}-\frac{56}{33}Q_{3,L}^{\mathrm{per}},
&&L\ge6.\label{eq:hierarchy-map-q5star}
\end{align}
Moreover, for the transfer-matrix convention of Eq.~\eqref{eq:transfer-matrix},
\begin{equation}\label{eq:q5-transfer-map}
\boxed{
\begin{aligned}
\widehat Q_L^{(5),\mathrm{per}}
&=16\overline T_{5,L},\\
Q_{5,L}^{\mathrm{boost},\mathrm{per}}
&=-\frac49 i\overline T_{5,L}-\frac{11}{36}Q_{3,L}^{\mathrm{per}},\\
Q_{5,L}^{\star,\mathrm{per}}
&=-\frac49 i\overline T_{5,L}-\frac{56}{33}Q_{3,L}^{\mathrm{per}}.
\end{aligned}
}
\end{equation}
The three fifth-charge relations hold on every unaliased periodic embedding \(L\ge6\). In particular,
\begin{equation}\label{eq:q5-transfer-normalized}
Q_{5,L}^{\mathrm{tr}}:=-\frac49 i\overline T_{5,L}
\end{equation}
is the transfer-normalized fifth hierarchy direction; \(Q_5^{\mathrm{boost}}\) and \(Q_5^\star\) differ from it by the displayed \(Q_3\) components.
The \(Q_4\) equalities are equalities of periodic translation sums. The raw boost density \(g\) and the centered transfer representative differ by the stated identity direction. The \(L=5\) conservation and commutativity statements used below are independent finite-ring checks and do not rely on the six-site normalization identity for \(Q_5\).
\end{proposition}

\begin{proof}
The scalar shift makes the one-site partial traces vanish and leaves all non-scalar commutators unchanged. The regular difference-form \(R\) matrix then gives \(B_{\mathrm{tr}}=B_{H^{(0)}}/2\); the ladder theorem and trace subtraction imply \(\widehat Q^{(n)}=2^{n-1}\overline T_n\). Periodic summation of the displayed \(q_4^\star\) formula fixes its \(H^{(0)}\) admixture, and one further boost step gives the \(Q_5^{\mathrm{boost}}\) relation. Equation~\eqref{eq:q5-hierarchy} then fixes the \(Q_5^\star\) coefficient. The scalar bookkeeping, finite-support step, and independent finite-ring transfer expansions are given in \ref{app:transfer-normalization-details}.
\end{proof}

The local Hilbert--Schmidt convention gives

\[
\left\langle
q_5^\star,
q_{3,5}^{A}
\right\rangle_{\mathrm{HS}}
=0.
\]

Here the inner product is evaluated in the aligned five-site window.

The remaining exact local properties are

\[
(q_5^\star)^\dagger=q_5^\star,
\qquad
\mathcal R_5q_5^\star\mathcal R_5=-q_5^\star,
\]

\[
[S^\alpha_{\mathrm{tot}},q_5^\star]=0,
\qquad
\operatorname{Tr}q_5^\star=0,
\qquad
\|q_5^\star\|_{\mathrm{HS}}^2=\frac{92310}{11}.
\]
The exact coefficient calculation and the subtraction of the embedded \(Q_3\) component are detailed in \ref{app:q5}.

\subsection{Local-divergence theorem}

Exact product-word orbit integration produces six-site integer currents \(\mathcal K_6^{\mathrm{boost}}\) and \(\mathcal K_6^\star\), where the subscript records the support width. In the Hermitian density normalization,

\begin{equation}\label{eq:q5boost-continuity}
\boxed{
[H,q_{5,j}^{\mathrm{boost}}]
=\frac{i}{216}
\left(
\mathcal K_{6,j}^{\mathrm{boost}}\otimes I
-I\otimes \mathcal K_{6,j}^{\mathrm{boost}}
\right),
}
\end{equation}
\begin{equation}\label{eq:q5can-continuity}
\boxed{
[H,q_{5,j}^{\star}]
=\frac{i}{396}
\left(
\mathcal K_{6,j}^{\star}\otimes I
-I\otimes \mathcal K_{6,j}^{\star}
\right).
}
\end{equation}
The same orbit integration applied to the seven-site local commutator density \(\mathcal M_7[A_3,A_5^\star]\) gives a six-site integer current \(K_{35}\):

\begin{equation}\label{eq:M7-divergence}
\mathcal M_7[A_3,A_5^\star]
=K_{35}\otimes I-I\otimes K_{35}.
\end{equation}
Its translation sum is \(-132[Q_{3,L}^{\mathrm{per}},Q_{5,L}^{\star,\mathrm{per}}]\). Thus all three local identities telescope for \(L\ge7\). Support aliasing at \(L=5,6\) is handled by independent exact integer matrices. The embedding convention, integer normalizations, currents, and residuals are given in \ref{app:orbit-integration}.

\begin{theorem}[Commuting hierarchy through five-site width]\label{thm:q5-periodic}

For every periodic chain of length \(L\ge5\),

\begin{equation}\label{eq:q5-conservation}
\boxed{
[H_L,Q_{5,L}^{\mathrm{boost},\mathrm{per}}]
=[H_L,Q_{5,L}^{\star,\mathrm{per}}]
=0,
}
\end{equation}
and

\begin{equation}\label{eq:q3q5-conservation}
\boxed{
[Q_{3,L}^{\mathrm{per}},Q_{5,L}^{\mathrm{boost},\mathrm{per}}]
=[Q_{3,L}^{\mathrm{per}},Q_{5,L}^{\star,\mathrm{per}}]
=0.
}
\end{equation}
Moreover,
\begin{equation}\label{eq:q4q5-conservation}
\boxed{
[Q_{4,L}^{\star,\mathrm{per}},Q_{5,L}^{\mathrm{boost},\mathrm{per}}]
=[Q_{4,L}^{\star,\mathrm{per}},Q_{5,L}^{\star,\mathrm{per}}]
=0,
\qquad L\ge5,
}
\end{equation}
and \([Q_{3,L}^{\mathrm{per}},Q_{4,L}^{\star,\mathrm{per}}]=0\) for every \(L\ge4\).
\end{theorem}

\begin{proof}
The Hamiltonian and \(Q_3\)--\(Q_5^\star\) statements follow from the local divergences, with small-ring completions. Equation~\eqref{eq:q5-hierarchy} then gives the \(Q_3\)--\(Q_5^{\mathrm{boost}}\) relation. For the remaining pairs at sufficiently large \(L\), Proposition~\ref{prop:hierarchy-normalization} verifies the normalization needed to place these charges in the boost hierarchy of \cite{R42} and identifies the corresponding transfer direction. Density divergences do not change periodic operators, and adding a lower charge already proved to commute with the other member leaves the relevant commutator unchanged.

The standing hypotheses of Corollary~2 of Ref.~\cite{R42} are met here. The local Hilbert space is finite dimensional, and \(h^{(0)}\) is Hermitian, translation invariant, nearest neighbor, and satisfies \(\operatorname{Tr}_1h^{(0)}=\operatorname{Tr}_2h^{(0)}=0\), as shown at the start of Proposition~\ref{prop:hierarchy-normalization}'s proof. Its boost recursion is exactly the recursion defining \(\widehat Q^{(n)}\) above; the scalar shift from \(H\) to \(H^{(0)}\) changes no commutator. At ring size six, Eqs.~\eqref{eq:q3-conservation} and \eqref{eq:hierarchy-map-q3} therefore give the required premise \([H_6^{(0)},\widehat Q_{6}^{(3),\mathrm{per}}]=0\), with the exact integer calculation providing an independent check.

The corollary states that the boost-hierarchy commutators vanish for \(N\ge4\) and integer levels \(2\le n,m\le\lfloor N/2\rfloor\). Hence it gives \([Q_3,Q_4]=0\) for \(L\ge8\) and \([Q_4,Q_5]=0\) for \(L\ge10\). Exact sparse integer calculations verify the two commutators for \(L=4,\ldots,9\) and \(L=5,\ldots,9\), respectively, while the corollary supplies their large-\(L\) continuation. Equations~\eqref{eq:q5-hierarchy} and \eqref{eq:q3q5-conservation} also imply \([Q_{5,L}^{\mathrm{boost},\mathrm{per}},Q_{5,L}^{\star,\mathrm{per}}]=0\). Equation~\eqref{eq:q4-representative-divergence} transfers every periodic \(Q_4^\star\) commutator to \(\widetilde Q_4\). Hence the named representatives generate pairwise commuting periodic charges on the stated rings. Equation~\eqref{eq:q5-transfer-map} independently fixes the transfer normalization of this same fifth hierarchy direction.
\end{proof}

\begin{corollary}[Nontriviality in the specified quotient and on small rings]\label{cor:q5-nontrivial}
The lower-charge-fixed density \(q_5^\star\) is nonzero in the specified translation-orbit quotient. On the periodic rings \(L=5,6\), its charge is nonzero and linearly independent of \(Q_{3,L}^{\mathrm{per}}\).
\end{corollary}

\begin{proof}
A pure local divergence has zero total coefficient in every finite-support translation orbit and is therefore annihilated by \(\mathcal L_5\), hence by \(\operatorname{Align}_{\mathrm{orb},5}^{(-)}\). By contrast, \(q_5^\star\) is a nonzero fixed point of the chosen alignment map, with local norm \(92310/11\). It is therefore nontrivial in this translation-orbit quotient.

The periodic statement can be checked directly without relying on the local quotient. Exact integer Gram matrices have rank two for the pairs \(\{Q_{3,L}^{\mathrm{per}},Q_{5,L}^{\star,\mathrm{per}}\}\) at \(L=5,6\). Hence \(Q_{5,L}^{\star,\mathrm{per}}\) is nonzero and linearly independent of \(Q_{3,L}^{\mathrm{per}}\) on both rings. The exact inner products are given in \ref{app:q5-small-ring-data}.
\end{proof}

\begin{corollary}[Low-order hierarchy independence]\label{cor:hierarchy-independence}
At \(L=5,6\), the sets \(\{H_L,I_L,Q_{3,L}^{\mathrm{per}},Q_{4,L}^{\star,\mathrm{per}}\}\) and \(\{H_L,Q_{3,L}^{\mathrm{per}},Q_{4,L}^{\star,\mathrm{per}},Q_{5,L}^{\star,\mathrm{per}}\}\) are linearly independent.
\end{corollary}

\begin{proof}
Exact Hilbert--Schmidt Gram matrices have rank four for both ordered sets at \(L=5,6\). In these rings, \(Q_4^\star\) is Hilbert--Schmidt orthogonal to \(H,I,Q_3\), and \(Q_5^\star\) is orthogonal to \(H,Q_3,Q_4^\star\). The nonzero determinants are recorded in \ref{app:q5-small-ring-data}.
\end{proof}

Corollaries~\ref{cor:q5-nontrivial} and \ref{cor:hierarchy-independence} record small-ring nontriviality and low-order linear independence. Table~\ref{tab:local-charges} collects the representative conventions and periodic statements used below.

\begin{table}[tbp]
\centering
\footnotesize
\caption{Local representatives and their periodic hierarchy content.}
\label{tab:local-charges}
\begin{tabularx}{\textwidth}{@{}>{\raggedright\arraybackslash}p{0.17\textwidth}*{3}{>{\raggedright\arraybackslash}X}@{}}
\toprule
Object & Local convention & Periodic content & Established range \\
\midrule
\(q_3\) (width 3, odd) & current/boost convention & canonical third boost level up to phase & \([H_L,Q_3]=0\), \(L\ge3\) \\
\(\widetilde q_4\) (width 4) & unaveraged \(g\)-based density & same periodic \(Q_4\) as \(q_4^\star\) & inherits the \(q_4^\star\) commutators for \(L\ge4\) \\
\(q_4^\star\) (width 4, even) & reflection average with local \(H,I\) directions fixed & fourth transfer/boost level, Eq.~\eqref{eq:hierarchy-map-q4} & commutes with \(H_L,Q_3\), \(L\ge4\) \\
\(Q_{5,L}^{\mathrm{tr}}\) & centered transfer normalization & fifth transfer direction & defined by the periodic transfer family; comparison to local representatives for \(L\ge6\) \\
\(q_5^{\mathrm{boost}}\) (width 5, odd) & ten-word divergence-reduced boost density & Eq.~\eqref{eq:q5-transfer-map} & commutes with \(H_L,Q_3,Q_4^\star,Q_5^\star\), \(L\ge5\) \\
\(q_5^{A}\) (width 5, odd) & orbit-aligned density gauge of \(q_5^{\mathrm{boost}}\) & same periodic charge as \(q_5^{\mathrm{boost}}\) & inherits all periodic commutators for \(L\ge5\) \\
\(q_5^\star\) (width 5, odd) & orbit alignment and stated local-HS lower-charge subtraction & convention-dependent direction in Eq.~\eqref{eq:q5-transfer-map} & commutes with \(H_L,Q_3,Q_4^\star,Q_5^{\mathrm{boost}}\), \(L\ge5\) \\
\bottomrule
\end{tabularx}
\end{table}

\section{Uniform MPO upper bounds}\label{sec:mpo-construction}

We now turn from conservation to representation cost. The uniform problem asks for one site tensor and fixed boundary vectors that work at every length, so the result is a reusable MPO template rather than a separate optimization for each \(L\). The represented object is the Hermitian truncated bulk sum of a chosen density; it is not automatically an open-chain conserved charge, for which explicit boundary counterterms would have to be supplied. This section gives upper bounds by factorizing one local density and packaging all its translates into a length-independent tensor. Section~\ref{sec:mpo-minimality} asks whether any smaller tensor can represent the same coefficient series.

\subsection{Exact tensor-train factorization}

Choose a one-site operator basis \(\{E_\mu\}_{\mu=0}^{8}\) and expand an \(r\)-site density as

\[
q^{(r)}
=\sum_{\mu_1,\ldots,\mu_r}
C_{\mu_1\cdots\mu_r}
E_{\mu_1}\otimes\cdots\otimes E_{\mu_r}.
\]

An exact tensor-train factorization is

\[
C_{\mu_1\cdots\mu_r}
=A_1^{\mu_1}A_2^{\mu_2}\cdots A_r^{\mu_r},
\]

with internal ranks

\[
(\rho_1,\ldots,\rho_{r-1}).
\]

These ranks are the exact matrix ranks of successive coefficient-tensor flattenings over the chosen coefficient field.

\subsection{Finite-state insertion}

Introduce an idle state, one active sector for every internal tensor-train bond, and a done state. The common one-site MPO tensor permits only the transitions

\[
\mathrm{idle}
\xrightarrow{A_1}
\rho_1
\xrightarrow{A_2}\cdots
\xrightarrow{A_r}
\mathrm{done},
\]

together with identity propagation in the idle and done states. Fixed boundary vectors select paths beginning in idle and ending in done. Every allowed path inserts exactly one copy of \(q^{(r)}\), while its starting site is arbitrary. Hence the contraction equals the truncated bulk sum \(\mathcal B_{r,L}[q]\).

The direct auxiliary dimension of this finite-state automaton (FSA) realization is

\begin{equation}\label{eq:fsa-dimension}
\boxed{
D_{\mathrm{FSA}}
=2+\sum_{k=1}^{r-1}\rho_k.
}
\end{equation}
This is the standard finite-state/weighted-automaton representation of a finite-support density \cite{R23,R24,R25,R26,R27,R35,R36,R37,R38,R39,R40}.

\subsection{\texorpdfstring{The \(q_4^\star\) upper bound}{The q4 star upper bound}}

Using the matrix-unit physical basis and exact rational arithmetic, the four-site coefficient tensor of \(216q_4^\star\) has tensor-train ranks

\begin{equation}\label{eq:q4-tt-ranks}
\boxed{
(\rho_1,\rho_2,\rho_3)=(9,26,9).
}
\end{equation}
The finite-state construction therefore gives

\[
D=1+9+26+9+1=46.
\]

The resulting rational \(46\times46\) auxiliary tensor, with left boundary normalization \(1/216\), represents \(\mathcal B_{4,L}[q_4^\star]\). Its finite-state structure is summarized in Fig.~\ref{fig:tt-fsa}; explicit tensor entries and boundary vectors are supplied in the Supplementary Computational Material.

\begin{figure}[tbp]
\centering
\includegraphics[width=\textwidth]{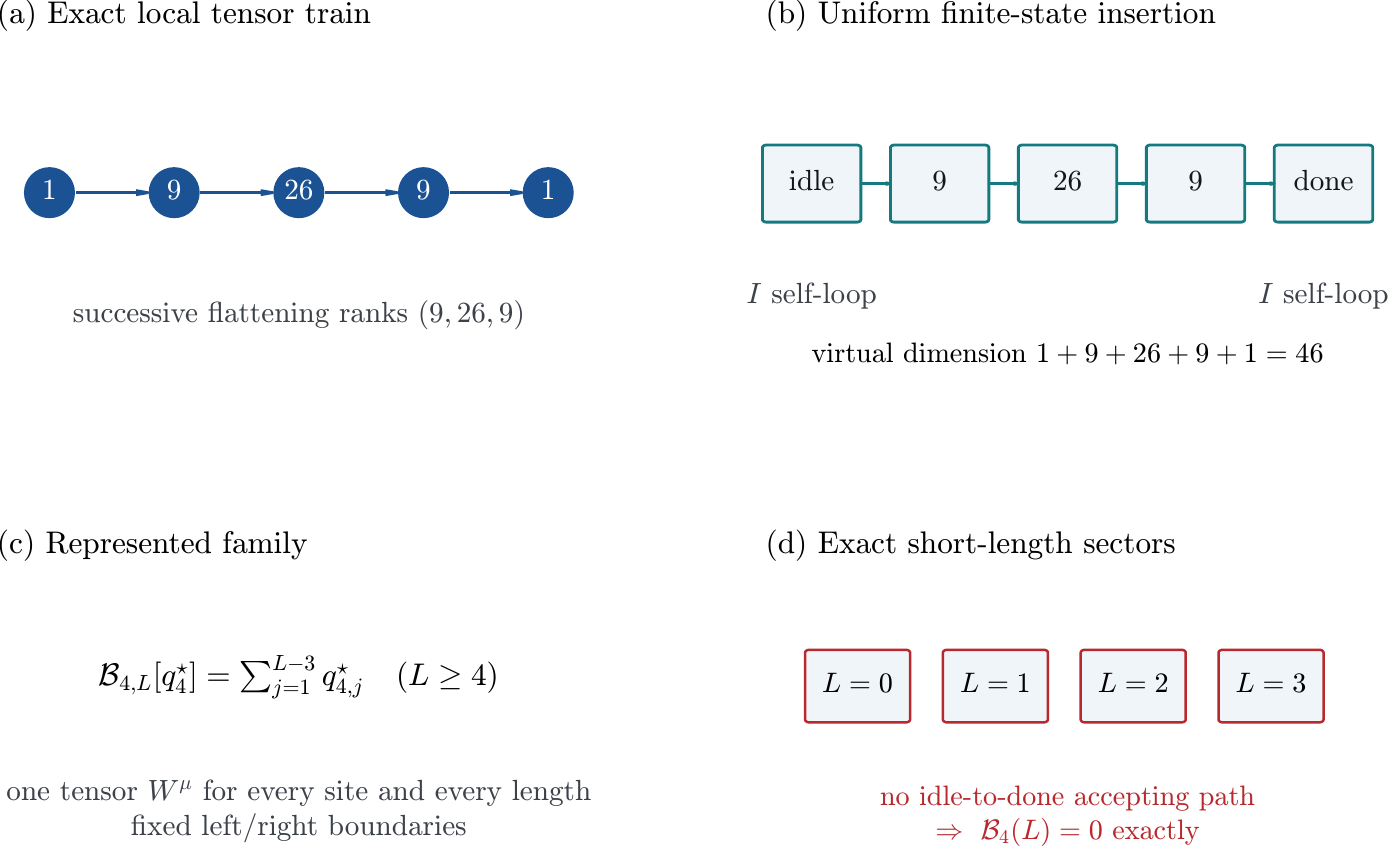}
\caption{Uniform fixed-boundary MPO construction. (a) The four-site density has successive-flattening ranks \((9,26,9)\). (b) The idle, three internal, and done sector dimensions sum to \(46\). (c) One tensor with fixed boundary vectors represents \(\mathcal B_{4,L}[q_4^\star]\) for every \(L\ge4\). (d) The absence of an idle-to-done path realizes the prescribed zero sectors for \(L<4\) exactly.}
\label{fig:tt-fsa}
\end{figure}

\subsection{Five-site representatives}

Exact rational flattenings of the compact boost density, its alignment-only gauge, and the lower-charge-fixed density give
\begin{equation}\label{eq:q5-tt-ranks}
\operatorname{TTRank}(q_5^{\mathrm{boost}})
=\operatorname{TTRank}(q_5^{A})
=\operatorname{TTRank}(q_5^\star)
=(9,33,33,9).
\end{equation}
The direct finite-state construction therefore gives the upper bound \(D_{\mathrm{FSA}}=86\) for all three. The equal ranks show that this selected fifth-order comparison has no TT-rank separation: \(q_5^{\mathrm{boost}}\sim_{\mathrm{div}}q_5^A\) is a pure gauge comparison, while the passage from \(q_5^A\) to \(q_5^\star\) also changes the hierarchy basis. Exact reproductions are described in \ref{app:q5-mpo-upper-bounds} and supplied with the Supplementary Computational Material.

\begin{corollary}[Fixed five-site full series]\label{cor:q5-full-series}
For each \(q\in\{q_5^{\mathrm{boost}},q_5^A,q_5^\star\}\), prescribe the
operator family to be zero for \(0\le L<5\) and to equal
\(\mathcal B_{5,L}[q]\) for \(L\ge5\).
Among uniform unblocked one-site MPOs with length-independent fixed
boundaries, its exact minimum auxiliary dimension is \(86\).
\end{corollary}
\begin{proof}
Lemma~\ref{lem:graded-full-series}, proved in
Sec.~\ref{sec:gauge-optima}, applies with width \(r=5\).
Combining it with Eq.~\eqref{eq:q5-tt-ranks} gives
\[
D_{\min}^{\mathrm{full}}[q]=2+9+33+33+9=86.
\]
The insertion tensor gives the upper bound. For the matching lower bound,
the graded Hankel construction selects local unfolding minors of sizes
\(1,9,33,33,9,1\), including the two end cuts. Their combined
\(86\times86\) Hankel matrix is nonsingular.
\end{proof}
The coefficients below length five are part of the full-series convention,
not physical short-ring charges. The dataset supplies the three exact
Hankel matrices and their prefix/suffix lists. Optimization over other
five-site gauges, free short-length completions, or periodic trace-MPOs is
not part of this corollary.

\section{Exact minimality and representative dependence for the four-site charge}\label{sec:mpo-minimality}

Because Theorem~\ref{thm:width4-scalar-quotient} leaves a unique nontrivial width-four scalar class, the comparison below isolates representative dependence within that class. The finite-state construction gives the upper bound \(46\). We prove optimality by specifying the represented coefficient family and MPO class, then exhibiting a rank-\(46\) Hankel witness. The definitions below also distinguish the full-series, tail-only, fixed-length, and periodic trace-MPO problems.

\subsection{Definition of the full series}

Fix the rational matrix-unit basis

\[
E_{3a+b}=|a\rangle\langle b|,
\qquad a,b\in\{0,1,2\},
\qquad \Sigma=\{0,\ldots,8\}.
\]

Here the matrix indices are labels for the ordered physical basis,

\[
|0\rangle_{\mathrm{idx}}\equiv|+1\rangle,\qquad
|1\rangle_{\mathrm{idx}}\equiv|0\rangle,\qquad
|2\rangle_{\mathrm{idx}}\equiv|-1\rangle,
\]

and are not magnetic quantum numbers. In this alphabet the physical identity is

\[
I=E_0+E_4+E_8.
\]

This is the basis used to construct the finite Hankel witnesses. Section~\ref{sec:basis-invariance} proves that the minimum is unchanged under any invertible one-site basis change.

Define the operator family

\begin{equation}\label{eq:full-family}
\mathfrak B_4(L)
=
\begin{cases}
0,&0\le L<4,\\[1mm]
\mathcal B_{4,L}[q_4^\star],&L\ge4.
\end{cases}
\end{equation}
Expanding every length in the product basis defines

\[
\mathfrak B_4(L)
=\sum_{w=(\mu_1\cdots\mu_L)\in\Sigma^L}
f_4(w)
E_{\mu_1}\otimes\cdots\otimes E_{\mu_L},
\]

and hence a coefficient function

\[
f_4:\Sigma^\ast\to\mathbb Q,
\qquad
f_4(w)=0\quad(|w|<4).
\]

The corresponding noncommutative formal series is

\[
\mathcal F_4=\sum_{w\in\Sigma^\ast}f_4(w)\,w.
\]

The zero sectors below length four are part of this formal coefficient-series definition rather than physical short-chain charges.

\subsection{Realization class}\label{sec:full-series-class}

We consider linear realizations

\begin{equation}\label{eq:linear-realization}
f_4(\mu_1\cdots\mu_L)
=\ell W^{\mu_1}\cdots W^{\mu_L}r
\end{equation}
subject to all of the following requirements:

\begin{enumerate}
\item one unblocked physical site per tensor;
\item one \(L\)-independent set of matrices \(W^\mu\);
\item \(L\)-independent boundary vectors \(\ell,r\);
\item ordinary virtual-space dimension;
\item agreement for every \(L\ge0\), including the prescribed zero sectors;
\item a coefficient field containing the density coefficients; for the displayed rational densities this may be \(\mathbb Q\), \(\mathbb R\), or \(\mathbb C\).
\end{enumerate}

The explicit finite-state tensor from Sec.~\ref{sec:mpo-construction} is a dimension-46 realization in this class.

For general Hermitian gauges we allow complex physical coefficients and work over \(\mathbb C\). The optimizing densities below are rational, so their minima are attained over \(\mathbb Q\) as well. For an irrational real affine parameter, the field must contain that parameter; a rational realization is not assumed.

Table~\ref{tab:realization-classes} separates this all-length problem from the three related realization classes used later.

\begin{table}[tbp]
\centering
\footnotesize
\caption{Realization class is part of every bond-dimension result.}
\label{tab:realization-classes}
\begin{tabularx}{\textwidth}{@{}>{\raggedright\arraybackslash}p{0.18\textwidth}*{3}{>{\raggedright\arraybackslash}X}@{}}
\toprule
Problem & Required coefficients & Tensors and boundaries & Result \\
\midrule
full coefficient series & every \(L\ge0\), including prescribed zeros for \(L<4\) & one unblocked one-site tensor and \(L\)-independent fixed boundaries & gauge-class minimum \(36\); reflection-even minimum \(44\); reference densities \(38,46\) \\
tail coefficient series & only \(L\ge4\); shorter words arbitrary & one unblocked one-site tensor and \(L\)-independent fixed boundaries & individual minima in Table~\ref{tab:gauge-optima}; gauge-global minimum open \\
one fixed \(L\ge4\) operator & coefficients of \(\mathcal B_{4,L}[q]\) & site-dependent fixed-boundary tensors & profiles differ; maxima agree at \(26\) for both when \(L\ge5\) \\
periodic trace-MPO & periodic family represented by an auxiliary trace & ordinary unblocked trace-MPO & \(D\ge9\) from \(L=4\); exact minimum open \\
\bottomrule
\end{tabularx}
\end{table}

\subsection{Hankel lower bound}

As a small example, a two-site density with coefficient matrix of rank \(\rho\) has a direct insertion automaton with idle, \(\rho\) active, and done sectors. Two prefixes can share an active state only when they induce the same linear response to every possible suffix. This is the general point: a cut through the auxiliary space must retain enough information about every prefix to predict all suffix coefficients, and Hankel rank counts the distinguishable responses exactly.

The Hankel matrix of the series is

\begin{equation}\label{eq:hankel-matrix}
\mathsf H_{u,v}=f_4(uv),
\qquad u,v\in\Sigma^\ast.
\end{equation}
For any \(D\)-dimensional realization,

\[
\mathsf H_{u,v}
=\bigl(\ell W^u\bigr)\bigl(W^v r\bigr),
\]

so every finite Hankel submatrix has rank at most \(D\). This is the noncommutative linear-realization/Hankel-rank principle established in the automata and sequential-system formulations of \cite{R35} and \cite{R36}. The historical attribution to Fliess \cite{R37} is documented by the review \cite{R41}, while \cite{R38} is a modern book-length background reference. The reachability and observability terminology follows the parallel control-theory formulation \cite{R39,R40}, while \cite{R27} explains its modern relevance for local MPOs.

There are 46 prefix words \(\{u_i\}\) and 46 suffix words \(\{v_j\}\), of lengths between zero and four, for which exact rational arithmetic gives

\begin{equation}\label{eq:hankel-witness}
\operatorname{rank}
\left[f_4(u_iv_j)\right]_{i,j=1}^{46}
=46.
\end{equation}
Equivalently, the associated reachability and observability matrices both have rank \(46\). Therefore every realization in the realization class defined above has \(D\ge46\). Together with the explicit upper bound:

\begin{theorem}[Full-series minimum]\label{thm:full-series-minimum}

Within the unblocked one-site, \(L\)-independent fixed-boundary realization class of Sec.~\ref{sec:full-series-class}, the coefficient function \(f_4\), equivalently the noncommutative formal series \(\mathcal F_4\), has minimum ordinary virtual-space dimension

\begin{equation}\label{eq:full-minimum}
\boxed{
D_{\min}^{\mathrm{full}}=46.
}
\end{equation}
Because the nonzero witness minor is rational, the same lower bound holds over \(\mathbb Q\), \(\mathbb R\), and \(\mathbb C\).
\end{theorem}

\begin{proof}
The finite-state insertion tensor of Sec.~\ref{sec:mpo-construction} gives the upper bound \(D\le46\). Equation~\eqref{eq:hankel-witness} gives a rank-46 Hankel submatrix, while every \(D\)-dimensional realization factors that submatrix through a \(D\)-dimensional auxiliary space. Hence \(D\ge46\). The witness entries are rational, so the same nonzero minor proves the lower bound after extending scalars to \(\mathbb R\) or \(\mathbb C\).
\end{proof}

Theorem~\ref{thm:full-series-minimum} concerns the specified truncated-bulk coefficient series of the reflection-even density \(q_4^\star\). Local divergences define different series, whose optimization at fixed width four is treated in Sec.~\ref{sec:gauge-optima}.

\subsection{Gauge sensitivity of full-series complexity}\label{sec:gauge-complexity}

Apply the same full-series definition and realization class to the unaveraged representative \(\widetilde q_4\) of Eq.~\eqref{eq:q4-tilde}. Exact successive flattenings give
\[
\operatorname{TTRank}(216\widetilde q_4)=(9,18,9),
\]
so the finite-state insertion construction has dimension \(1+9+18+9+1=38\). The explicit rational realization has cumulative reachable and observable dimensions
\[
(1,10,28,37,38).
\]
Selected prefix and suffix words give an exact rank-38 Hankel submatrix, providing the matching lower bound.

\begin{theorem}[Representative-dependent full-series complexity]\label{thm:gauge-complexity}
For the realization class of Sec.~\ref{sec:full-series-class},
\begin{equation}\label{eq:gauge-complexity}
\boxed{
D_{\min}^{\mathrm{full}}[\widetilde q_4]=38,
\qquad
D_{\min}^{\mathrm{full}}[q_4^\star]=46.
}
\end{equation}
Nevertheless, Eq.~\eqref{eq:q4-representative-divergence} implies
\(Q_L^{\mathrm{per}}[\widetilde q_4]=Q_L^{\mathrm{per}}[q_4^\star]\) for every \(L\ge4\).
\end{theorem}

\begin{proof}
The exact TT insertion constructions give dimensions \(38\) and \(46\). Corresponding finite Hankel submatrices have ranks \(38\) and \(46\), which are matching lower bounds by the same factorization argument as in Theorem~\ref{thm:full-series-minimum}. Periodic equality follows by summing the local divergence in Eq.~\eqref{eq:q4-representative-divergence} around the ring.
\end{proof}

Thus reflection symmetrization leaves the periodic conserved operator unchanged but modifies the truncated-bulk coefficient series. For these reference densities it changes the TT middle rank from \(18\) to \(26\) and the minimal full-series dimension from \(38\) to \(46\). Neither representative minimizes the complete fixed-width gauge problem considered next.

\subsection{Exact optimization over four-site density gauges}\label{sec:gauge-optima}

We now minimize the full-series dimension over all three-site Hermitian
divergence currents. Fix \(G_0=216q_4^\star\) and define
\begin{equation}\label{eq:four-site-gauge-class}
\begin{aligned}
\mathcal G_4&=\{G_0+\Delta_3 C:C=C^\dagger\in
\operatorname{End}[(\mathbb C^3)^{\otimes3}]\},\\
\Delta_3 C&=C\otimes I-I\otimes C,\\
\mathcal G_4^+&=\{G\in\mathcal G_4:\mathcal R_4G\mathcal R_4=G\}.
\end{aligned}
\end{equation}
The identity component of \(C\) has zero divergence. Its other \(27^2-1=728\)
real Hermitian directions are all allowed, with no SU(2) restriction. Dividing
\(G\) by \(216\) restores the physical normalization and changes no rank.
Every member generates the same periodic charge for \(L\ge4\).
The class fixes support width, charge scale, and lower-charge components.
On an open interval the divergence leaves two endpoint terms:
\[
\mathcal B_{4,L}[G]-\mathcal B_{4,L}[G_0]
=\tau_1(C)-\tau_{L-2}(C),\qquad L\ge4,
\]
where the translated currents are padded with identities to length \(L\).
Thus the optimization changes how one periodic charge is represented locally,
through its boundary data in the truncated bulk sum.

\begin{lemma}[Graded full-series rank]\label{lem:graded-full-series}
Let a nonzero width-\(r\) density have local TT ranks
\(\rho_1,\ldots,\rho_{r-1}\). For its translated truncated-bulk series, with
coefficients prescribed to vanish below length \(r\),
\begin{equation}\label{eq:graded-full-series}
D_{\min}^{\mathrm{full}}=2+\sum_{j=1}^{r-1}\rho_j.
\end{equation}
\end{lemma}
\begin{proof}
The TT insertion construction gives the upper bound. For the lower bound,
let \(f\) be the full-series coefficient function and set \(\rho_0=\rho_r=1\).
For every cut \(j\), choose a nonsingular \(\rho_j\)-square minor of the local
unfolding, including nonzero one-entry minors at the two end cuts. Its row and
column words have lengths \(j\) and \(r-j\). Collect these prefixes and suffixes
in a single Hankel submatrix, ordering both groups by increasing cut index.
For prefixes \(u\) from group \(j\) and suffixes \(v\) from group \(j'\),
\[
|uv|=j+(r-j')=r+j-j'.
\]
The Hankel entry \(f(uv)\) therefore vanishes when \(j<j'\). On the diagonal,
\(|uv|=r\), where the truncated bulk sum contains exactly one copy of the
density. Each diagonal block is consequently the selected local unfolding
minor. The resulting block lower-triangular matrix is nonsingular and has
size \(\sum_{j=0}^r\rho_j\). Every dimension-\(D\) realization factors this
matrix through its auxiliary space, so \(D\ge\sum_{j=0}^r\rho_j\), matching
the insertion bound.
\end{proof}
The prescribed short-length zeros make the Hankel matrix block triangular.
For width four, the lemma reduces the gauge optimization to minimizing
\(\rho_1+\rho_2+\rho_3\).

Two simple densities attain the gauge-class bounds. With normalized partial
traces of the fixed starting density, put
\begin{equation}\label{eq:gauge-partial-traces}
E=\frac13\operatorname{Tr}_1G_0,\qquad
T=\frac19\operatorname{Tr}_{1,2}G_0
 =36I_9-21d-27d^2,\qquad d=\boldsymbol S_1\cdot\boldsymbol S_2.
\end{equation}
Define an oriented density and a reflection-even density by
\begin{equation}\label{eq:gauge-optimizing-densities}
\begin{aligned}
G_{\mathrm L}&=G_0+\Delta_3(E+T\otimes I),\\
G_{\mathrm E}&=G_0+\Delta_3(I\otimes T-T\otimes I),\qquad
q_{\mathrm L}=G_{\mathrm L}/216,\quad q_{\mathrm E}=G_{\mathrm E}/216.
\end{aligned}
\end{equation}
These choices have a direct interpretation. The normalized three-site partial
trace of \(G_0\) vanishes, so \(\operatorname{Tr}_1T=0\). For
\(C_{\mathrm L}=E+T\otimes I\), we have
\(\frac13\operatorname{Tr}_1C_{\mathrm L}=T\), and hence
\[
\frac13\operatorname{Tr}_1G_{\mathrm L}
=E+T\otimes I-C_{\mathrm L}=0.
\]
The left identity sector has been removed: each translation orbit is placed
at the left edge. The second construction can be written as
\[
G_{\mathrm E}=G_0-T\otimes I^{\otimes2}
                 +2I\otimes T\otimes I-I^{\otimes2}\otimes T.
\]
It moves adjacent-pair contributions from the two edges to the center of the
four-site window and is reflection even. Both densities are Hermitian SU(2)
scalars. Their exact TT ranks are \((8,17,9)\) and \((9,24,9)\), respectively,
giving the candidate dimensions \(36\) and \(44\). The next proof establishes
lower bounds for every density in the corresponding classes.

\begin{theorem}[Four-site gauge-class minima]\label{thm:gauge-optima}
In the full-series realization class of Sec.~\ref{sec:full-series-class},
\begin{equation}\label{eq:gauge-class-minima}
\boxed{
\min_{G\in\mathcal G_4}D_{\min}^{\mathrm{full}}[G/216]=36,
\qquad
\min_{G\in\mathcal G_4^+}D_{\min}^{\mathrm{full}}[G/216]=44.
}
\end{equation}
The densities \(q_{\mathrm L}\) and \(q_{\mathrm E}\) attain the respective minima.
\end{theorem}
\begin{proof}
The constructions above and Lemma~\ref{lem:graded-full-series} give the upper
bounds. We prove the lower bounds in four steps.
The first two steps apply to an arbitrary Hermitian current. The third
excludes the only rank pattern that could lower the dimension below \(36\);
the fourth imposes reflection symmetry and yields \(44\).

\emph{1. Coefficients fixed by the periodic charge.}
Expand an arbitrary \(G\in\mathcal G_4\) in the orthogonal Hermitian one-site
basis \(H_0=I,H_1,\ldots,H_8\) specified in \ref{app:gauge-optima}:
\[
G=\sum_{a,b,c,d=0}^8g_{abcd}H_a\otimes H_b\otimes H_c\otimes H_d.
\]
Hermiticity makes every \(g_{abcd}\) real. In the following, an active index
runs from \(1\) to \(8\). A current with coefficients \(c_{abc}\) changes
\(g_{abcd}\) by \(c_{abc}\delta_{d0}-\delta_{a0}c_{bcd}\). It therefore
preserves the tensors
\[
\begin{aligned}
A_{abcd}&=g_{abcd},& B_{abc}&=g_{abc0}+g_{0abc},\\
M_{ab}&=g_{a0b0}+g_{0a0b},&
N_{ab}&=g_{ab00}+g_{0ab0}+g_{00ab}.
\end{aligned}
\]
These are the four-active-site coefficients and the orbit sums for three
consecutive active sites, separated pairs, and adjacent pairs. Substitution
of \(G_0\) gives
\[
M=\operatorname{diag}(-21,-21,-42,-18,-18,-9,-9,-6).
\]
All three unfoldings of \(A\) have rank \(8\). Coefficients with active first
and last indices and an interior zero vanish; the one-active-site orbit sums
and \(g_{0000}\) also vanish. These fixed properties, together with the short
contractions below, are the finite algebraic inputs to the proof.

\emph{2. A lower bound valid for every gauge.}
Group the middle unfolding by whether its first and last indices are zero:
\[
\begin{pmatrix}X&Y_{\mathrm{full}}\\Z_{\mathrm{full}}&K\end{pmatrix}.
\]
Here \(K\) has active first and last indices. Its only nonzero part is the
middle unfolding of \(A\), so \(\operatorname{rank}K=8\). Reducing \(K\)
to rank normal form separates eight pivots from its left and right kernels.
The residual matrix has a zero lower-right block. Its rank is at least the
sum of the ranks of its two off-diagonal blocks, as detailed in
Eq.~\eqref{eq:gauge-bordered-rank}. The zero interior-index columns and rows
of \(K\) expose, respectively,
\[
Y_{ab}=g_{0a0b},\qquad Z_{ab}=g_{a0b0},\qquad Y+Z=M.
\]
It follows that
\[
\rho_2\ge8+\operatorname{rank}Y+\operatorname{rank}Z
\ge8+\operatorname{rank}M=16.
\]
The outer unfoldings contain the corresponding rank-eight unfoldings of
\(A\), so each outer rank is either \(8\) or \(9\). If both are \(9\),
the full-series dimension is already at least \(2+9+16+9=36\).
It remains to consider a possible drop of an outer rank to \(8\).

\emph{3. An outer rank of eight forces middle rank seventeen.}
Suppose \(\rho_1=8\). The eight active-first-index rows are independent
because their restriction to \(A\) has rank eight. The identity row must
therefore be their unique linear combination:
\[
g_{0bcd}=\sum_{a=1}^8v_a g_{abcd},\qquad b,c,d=0,\ldots,8,
\quad v\in\mathbb R^8.
\]
This relation is necessary for every candidate with \(\rho_1=8\).
The vector \(v\) records a row dependence of that candidate, so no
restriction to a chosen family of currents has been introduced.
For a lower bound, the necessary relations suffice: every competing
density must satisfy them.
Applying this relation to \(g_{0a0b}\) and using the fixed interior zeros gives
\(Y=0\), hence \(Z=M\). The same row relation gives
\(g_{00ab}=g_{000a}=0\). The orbit sums now determine the remaining
coefficients in terms of \(v\):
\[
\begin{aligned}
\ell_{abc}:=g_{abc0}&=B_{abc}-\sum_dv_dA_{dabc},\\
P_{ab}:=g_{ab00}&=N_{ab}-\sum_dv_d\ell_{dab},\\
u_a:=g_{a000},\qquad u&=-(P+M)^{\mathsf T}v,\qquad p:=v^{\mathsf T}u=0.
\end{aligned}
\]
The four one-site placements have coefficients \(u_a\),
\((P^{\mathsf T}v)_a\), \((M^{\mathsf T}v)_a\), and zero. Their vanishing
sum gives the formula for \(u\). The final condition follows from the
identity row at \((b,c,d)=(0,0,0)\), since
\(g_{0000}=0\). Thus \(p=0\) is a compatibility condition on a candidate,
not an additional choice of gauge.

Delete the dependent first-index-zero rows of the middle unfolding and its
zero columns with third index zero and fourth index active. The remaining
rows \((a,0),(a,b)\) and columns \((c,0),(c,d),(0,0)\) give
\[
\begin{pmatrix}M&0&u\\\ell&A&\operatorname{vec}P\end{pmatrix}.
\]
In this matrix \(\ell_{(a,b),c}=\ell_{abc}\) and
\(A_{(a,b),(c,d)}=A_{abcd}\). Elimination with the invertible block \(M\)
therefore gives
\[
\rho_2=8+\operatorname{rank}[A,\operatorname{vec}V],\qquad
V_{ab}=P_{ab}-\sum_c\ell_{abc}(M^{-1}u)_c.
\]
Since \(A\) has rank eight, a middle rank of \(16\) would require
\(\operatorname{vec}V\in\operatorname{col}A\).

We exclude this possibility by a scalar contraction with the weight
\[
\Omega=\operatorname{diag}(66,66,33,17,17,34,34,51).
\]
The weight follows from a linear system involving only \(A,B,M\).
For diagonal entries \(w_a=\Omega_{aa}\), the required conditions are
\[
\sum_a w_aB_{daa}=\sum_a w_aB_{aac}=0,\qquad
\sum_a w_aA_{eaac}=-9M_{ec}.
\]
In the stated basis this rational linear system has rank eight and gives
the displayed weights. The normalization \(9\) fixes their overall
scale. Their role is to cancel the variable-dependent terms in the
trace of the reduced column; the remaining nonzero constant will
exclude rank sixteen.
Direct expansion of \(G_0\) gives the finite identities
\[
\begin{aligned}
A_{aacd}&=A_{abcc}=0,&
\sum_a\Omega_{aa}B_{daa}&=\sum_a\Omega_{aa}B_{aac}=0,\\
\sum_a\Omega_{aa}A_{eaac}&=-9M_{ec},&
\operatorname{Tr}(\Omega N)&=-3780.
\end{aligned}
\]
Substitution of \(\ell=B-vA\) first yields
\[
\sum_a\Omega_{aa}\ell_{daa}=0,\qquad
\sum_a\Omega_{aa}\ell_{aac}=9(Mv)_c.
\]
The first equality gives \(\operatorname{Tr}(\Omega P)=-3780\); the second
then evaluates the trace of \(V\):
\[
\begin{aligned}
\operatorname{Tr}(\Omega V)
&=\operatorname{Tr}(\Omega P)
  -\sum_c\left(\sum_a\Omega_{aa}\ell_{aac}\right)(M^{-1}u)_c\\
&=-3780-9(Mv)^{\mathsf T}M^{-1}u
 =-3780-9p=-3780.
\end{aligned}
\]
Every vector in \(\operatorname{col}A\) has zero entries in its diagonal
rows \((a,a)\), because \(A_{aacd}=0\). If \(\operatorname{vec}V\) belonged
to that space, its weighted trace would be zero, contradicting the displayed
value. Appending \(\operatorname{vec}V\) therefore raises the rank of \(A\)
by one, so \(\rho_2=17\).

Reflection gives the corresponding result when \(\rho_3=8\). The two outer
ranks cannot both be eight: the left condition forces \(Y=0\), and the right
condition forces \(Z=0\), contradicting \(Y+Z=M\). Hence all possible rank
patterns obey
\[
\begin{array}{c|c|c}
(\rho_1,\rho_3)&\rho_2&D_{\min}^{\mathrm{full}}\\ \hline
(9,9)&\ge16&\ge36\\
(8,9)\text{ or }(9,8)&17&36.
\end{array}
\]
In particular, the only remaining way to obtain dimension \(35\), namely
one outer rank eight and middle rank sixteen, has been excluded.

\emph{4. The reflection-even constraint.}
Reflection sends \(Y\) to \(Z^{\mathsf T}\), so an even density obeys
\(Z=Y^{\mathsf T}\) and \(Y+Y^{\mathsf T}=M\). If the real matrix \(Y\)
had a nonzero real null vector \(x\), then
\[
x^{\mathsf T}Mx=x^{\mathsf T}(Y+Y^{\mathsf T})x=0,
\]
contradicting the negative definiteness of \(M\). Both \(Y\) and \(Z\)
therefore have rank eight, giving \(\rho_2\ge8+8+8=24\). Neither outer
rank can be eight, since that would force \(Y\) or \(Z\) to vanish.
Thus \(D\ge2+9+24+9=44\). Together with the constructions above, these
bounds prove both minima over the stated Hermitian classes.
\end{proof}

\begin{table}[!b]
\centering
\caption{Four gauges of the same periodic \(Q_4\). Full-series and tail values
refer to each displayed density. The full-series values \(36\) and \(44\)
also attain the minima over \(\mathcal G_4\) and \(\mathcal G_4^+\), respectively.}
\label{tab:gauge-optima}
\begin{tabular}{@{}lcccc@{}}
\toprule
Density & Reflection even & TT ranks & \(D_{\min}^{\mathrm{full}}\) & \(D_{\min}^{\mathrm{tail}}\)\\
\midrule
\(\widetilde q_4\) & no & \((9,18,9)\) & 38 & 38\\
\(q_4^\star\) & yes & \((9,26,9)\) & 46 & 46\\
\(q_{\mathrm L}\) & no & \((8,17,9)\) & 36 & 36\\
\(q_{\mathrm E}\) & yes & \((9,24,9)\) & 44 & 44\\
\bottomrule
\end{tabular}
\end{table}

The extra eight auxiliary states quantify the cost of requiring a
reflection-even local representative of the same reflection-even periodic
charge.

The individual tail equalities for the two
optimizers follow from the restriction-dimension argument in
\ref{app:gauge-tail}. The complete basis and coefficient contractions used
above are recorded in \ref{app:gauge-optima}; the reproduction package supplies
the exact tensors and executable checks. Gauge optimization for the tail-only
problem, larger density widths, and periodic trace-MPOs remains open.

\subsection{An affine family of density gauges}\label{sec:affine-gauge-family}
Define
\begin{equation}\label{eq:affine-gauge-family}
q_4(\lambda)=(1-\lambda)\widetilde q_4+\lambda q_4^\star,
\qquad \lambda\in\mathbb R.
\end{equation}
Equation~\eqref{eq:q4-representative-divergence} implies that every member generates the same periodic \(Q_4\) for \(L\ge4\). Moreover,
\[
\mathcal R_4q_4(\lambda)\mathcal R_4=q_4(2-\lambda).
\]
The reflection-even density is at \(\lambda=1\); the densities at \(0\) and \(2\) are related by reflection.

\begin{theorem}[Full-series complexity along the affine gauge family]\label{thm:affine-gauge-family}
In the realization class of Sec.~\ref{sec:full-series-class},
\begin{equation}\label{eq:affine-gauge-minimum}
D_{\min}^{\mathrm{full}}[q_4(\lambda)]=
\begin{cases}
38,&\lambda\in\{0,2\},\\
46,&\lambda\in\mathbb R\setminus\{0,2\}.
\end{cases}
\end{equation}
The corresponding local TT ranks are \((9,18,9)\) at \(0,2\) and \((9,26,9)\) elsewhere.
\end{theorem}

\begin{proof}
The three unfoldings of \(216q_4(\lambda)\) are polynomial matrices. Exact elimination over \(\mathbb Q(\lambda)\) gives ranks \((9,26,9)\). Hence every higher-order minor vanishes identically, and specialization to any parameter has ranks at most these values. The insertion construction therefore gives the uniform bound \(D\le46\) at every parameter.

Two fixed \(46\times46\) Hankel submatrices have nonzero determinant polynomials whose monic greatest common divisor is \(\lambda^8(\lambda-2)^8\). Their explicit factorizations and word lists are described in \ref{app:affine-hankel}. At any \(\lambda\ne0,2\), at least one determinant is nonzero, proving \(D\ge46\). Since the insertion bound is \(2+r_1+r_2+r_3\), this equality also forces the three local ranks to be \((9,26,9)\). At \(0,2\), exact ranks \((9,18,9)\) give \(D\le38\), while rank-38 Hankel witnesses give the matching lower bound.
\end{proof}

Thus any sufficiently small nonzero displacement from either low-dimension point along this line raises the exact full-series minimum from \(38\) to \(46\). The periodic charge stays fixed throughout. The line contains neither the dimension-36 optimum nor the reflection-even dimension-44 optimum of Theorem~\ref{thm:gauge-optima}. It illustrates rank changes within a restricted family. The tail-only theorem below concerns the two reference densities.

\subsection{Invariance under a one-site basis change}\label{sec:basis-invariance}

The values in Theorems~\ref{thm:full-series-minimum} and \ref{thm:gauge-complexity} are invariant under invertible changes of the one-site coordinate basis.

\begin{lemma}[Basis invariance]\label{lem:basis-invariance}

Fix \(\mathbb K\in\{\mathbb Q,\mathbb R,\mathbb C\}\). Let \(\widetilde E_\nu=\sum_\mu U_{\nu\mu}E_\mu\), where \(U\in\mathrm{GL}(9,\mathbb K)\). Then the minimum ordinary virtual dimension over \(\mathbb K\) of the corresponding full coefficient series is unchanged.
\end{lemma}

\begin{proof} Since \(U\) is invertible over \(\mathbb K\), one has \(E_\mu=\sum_\nu(U^{-1})_{\mu\nu}\widetilde E_\nu\). Given a realization \(\{W^\mu\}\) over \(\mathbb K\), the coefficient series in the transformed basis is therefore realized by
\[
\widetilde W^\nu=\sum_\mu(U^{-1})_{\mu\nu}W^\mu,
\]
without changing the common auxiliary space. The inverse basis change gives the converse inequality, and all coefficients remain in \(\mathbb K\). Thus the two minimum dimensions over \(\mathbb K\) coincide.
\end{proof}

\subsection{Tail-only realization}

For a density \(q\in\{q_4^\star,\widetilde q_4\}\), a tail competitor agrees with the corresponding truncated-bulk series on every word of length at least four. Its shorter-word coefficients are arbitrary. The transition matrices and boundaries constructed above remain feasible tail realizations, so they give upper bounds \(46\) and \(38\), respectively. Allowing arbitrary shorter-word coefficients cannot reduce either dimension.

\begin{theorem}[Tail-completion rigidity]\label{thm:tail-rigidity}
In the same unblocked one-site, \(L\)-independent fixed-boundary class,
\begin{equation}\label{eq:tail-minima}
\boxed{
D_{\min}^{\mathrm{tail}}[q_4^\star]=46,
\qquad
D_{\min}^{\mathrm{tail}}[\widetilde q_4]=38.
}
\end{equation}
Thus the representative-dependent separation persists without prescribing any coefficient below length four.
\end{theorem}

The proof uses a dimension argument. Length-four rows already force a \(36\)-dimensional space; restricting them to long suffixes retains only \(26\) dimensions. The remaining fixed restricted rows must therefore supply the missing directions, forcing the totals \(46\) and \(38\).

\begin{proof}
A tail competitor of infinite Hankel rank cannot improve either finite upper bound. It therefore suffices to consider a finite-dimensional Hankel row space \(V\) and set \(R:=\dim V\). All rows indexed by words \(u\) of length four are completely fixed, because \(|uv|\ge4\) for every suffix \(v\). Their span \(S\subseteq V\) has exact dimension \(36\) for both representatives. Let \(\rho\) restrict a row function to columns indexed by words of length at least four. Exact rational elimination gives
\[
\dim\rho(S)=26.
\]
The restrictions of all rows with \(|u|\le4\) are likewise fixed by the tail data. Their spans have dimensions \(36\) for \(q_4^\star\) and \(28\) for \(\widetilde q_4\).

Choose a vector-space complement \(V=S\oplus W\), so \(\dim W=R-36\). Since \(\rho(V)\subseteq\rho(S)+\rho(W)\),
\[
\dim\rho(V)\le26+(R-36).
\]
The fixed restricted rows lie in \(\rho(V)\). Therefore
\[
36\le26+(R-36)\quad\Longrightarrow\quad R\ge46
\]
for \(q_4^\star\), whereas
\[
28\le26+(R-36)\quad\Longrightarrow\quad R\ge38
\]
for \(\widetilde q_4\). The explicit \(46\)- and \(38\)-dimensional realizations attain these lower bounds. Exact rational data for the restricted row spaces and quotient dimensions are listed in \ref{app:crosswalk}.
\end{proof}

\subsection{\texorpdfstring{Fixed-length realizations for every \(L\ge4\)}{Fixed-length realizations for every L >= 4}}

With one-site basis dependence excluded by Lemma~\ref{lem:basis-invariance}, consider a single fixed chain with site-dependent fixed-boundary tensors. The minimum bond dimension at each cut equals the corresponding exact operator-Schmidt rank \cite{R28,R29}. Indeed, an MPO bond of dimension \(D_j\) yields a sum of at most \(D_j\) product operators across cut \(j\), so \(D_j\) is bounded below by that cut's operator-Schmidt rank. Conversely, successive exact Schmidt factorizations attain all these cut ranks simultaneously.

For the explicit 46-dimensional realization, define the exact-length row and column spaces
\[
\mathsf{Reach}_n=\operatorname{span}\{\ell W^w:|w|=n\},
\qquad
\mathsf{Obs}_n=\operatorname{span}\{W^w r:|w|=n\}.
\]

\begin{lemma}[Exact-length stabilization and all-length cut ranks]\label{lem:fixed-all-lengths}
The spaces \(\mathsf{Reach}_n\) and \(\mathsf{Obs}_n\) have dimensions
\[
(1,9,26,35,36,36)\qquad(n=0,\ldots,5),
\]
and satisfy \(\mathsf{Reach}_n=\mathsf{Reach}_4\) and \(\mathsf{Obs}_n=\mathsf{Obs}_4\) for every \(n\ge4\). For every \(L\ge4\), the coefficient matrix across cut \(j\) has rank \(9\) when \(j=1\) or \(L-1\), and rank \(26\) at every interior cut.
\end{lemma}

\begin{proof}
Exact rational elimination gives the displayed dimensions and, more strongly,
\[
\dim(\mathsf{Reach}_4+\mathsf{Reach}_5)=36,
\qquad
\dim(\mathsf{Obs}_4+\mathsf{Obs}_5)=36.
\]
Since each summand already has dimension \(36\), these equalities imply \(\mathsf{Reach}_5=\mathsf{Reach}_4\) and \(\mathsf{Obs}_5=\mathsf{Obs}_4\). The transition recursions
\[
\mathsf{Reach}_{n+1}=\operatorname{span}_{\mu}\mathsf{Reach}_nW^\mu,
\qquad
\mathsf{Obs}_{n+1}=\operatorname{span}_{\mu}W^\mu\mathsf{Obs}_n
\]
then prove invariance by induction for all \(n\ge4\).

Across a length-\(L\) cut after site \(j\), the coefficient matrix factors as the pairing of \(\mathsf{Reach}_j\) with \(\mathsf{Obs}_{L-j}\). Stabilization reduces every \(L\ge4\) cut to the finite cases in which both length arguments are capped at four. Exact elimination of those pairings gives rank \(9\) for classes \((1,3)\) and \((1,4)\), rank \(26\) for every class \((a,b)\) with \(a,b\ge2\), and the reflected cases identically. These are respectively the two edge cuts and all interior cuts. The exact MPO tensors and word bases needed to reconstruct these finite pairings are supplied in the Supplementary Computational Material.
\end{proof}

Lemma~\ref{lem:fixed-all-lengths} gives

\begin{equation}\label{eq:fixed-schmidt-profile}
\operatorname{SchmidtProfile}
\bigl(\mathcal B_{4,L}[q_4^\star]\bigr)
=\left(9,\underbrace{26,\ldots,26}_{L-3},9\right),
\qquad L\ge4.
\end{equation}
Hence

\begin{equation}\label{eq:fixed-minimum}
\boxed{
\min_{\text{site-dependent fixed-}L\text{ MPOs}}
\max_jD_j=26
\qquad(L\ge4).
}
\end{equation}
The divergence-equivalent representative \(\widetilde q_4\) has a different fixed-length profile.

\begin{proposition}[Fixed-length representative comparison]\label{prop:fixed-representative-comparison}
For \(\widetilde q_4\),
\begin{equation}\label{eq:fixed-schmidt-profile-tilde}
\operatorname{SchmidtProfile}\bigl(\mathcal B_{4,L}[\widetilde q_4]\bigr)
=
\begin{cases}
(9,18,9),&L=4,\\
\left(9,\underbrace{26,\ldots,26}_{L-4},18,9\right),&L\ge5.
\end{cases}
\end{equation}
Consequently, its optimal maximum fixed-length bond dimension is \(18\) at \(L=4\) and \(26\) for every \(L\ge5\). Thus the two representatives have different complete profiles for every \(L\ge4\), while their optimal maximum dimensions agree for all \(L\ge5\).
\end{proposition}

\begin{proof}
For the exact 38-dimensional realization of \(\widetilde q_4\), the row-space dimensions at lengths zero through five are \((1,9,26,35,36,36)\), the column-space dimensions are \((1,9,18,27,28,28)\), and the length-four/five union ranks are \(36\) and \(28\). The same transition-space induction as in Lemma~\ref{lem:fixed-all-lengths} reduces every cut to a \(4\times4\) table of length classes. Exact rational elimination gives the edge sequence \(9\), the right-near-edge rank \(18\), and rank \(26\) at every remaining interior cut. Reconstruction of the transition spaces and these finite pairings determines the complete profiles for every \(L\ge4\).
\end{proof}

\subsection{A periodic trace-MPO lower bound}

\begin{proposition}[Periodic trace-MPO obstruction]\label{prop:periodic-trace-bound}
Any ordinary unblocked periodic trace-MPO representing \(Q_{4,4}^{\star,\mathrm{per}}\) has auxiliary dimension
\[
D\ge9.
\]
Consequently, the same lower bound applies to any uniform periodic trace-MPO family representing \(Q_{4,L}^{\star,\mathrm{per}}\) for every \(L\ge4\).
\end{proposition}

\begin{proof}
Cutting a trace-MPO across a bipartition opens two virtual bonds. A bond-\(D\) tensor therefore expresses the operator as a sum of at most \(D^2\) left--right product operators, so its operator-Schmidt rank \(\sigma\) obeys \(\sigma\le D^2\). Exact rational reshaping of the four-site periodic charge gives
\[
\operatorname{SchmidtProfile}
\bigl(Q_{4,4}^{\star,\mathrm{per}}\bigr)=(9,74,9).
\]
At the middle cut, \(74\le D^2\), hence \(D\ge\lceil\sqrt{74}\rceil=9\). The exact rank calculation is included in \ref{app:crosswalk}.
\end{proof}

\begin{figure}[!t]
\centering
\includegraphics[width=\textwidth]{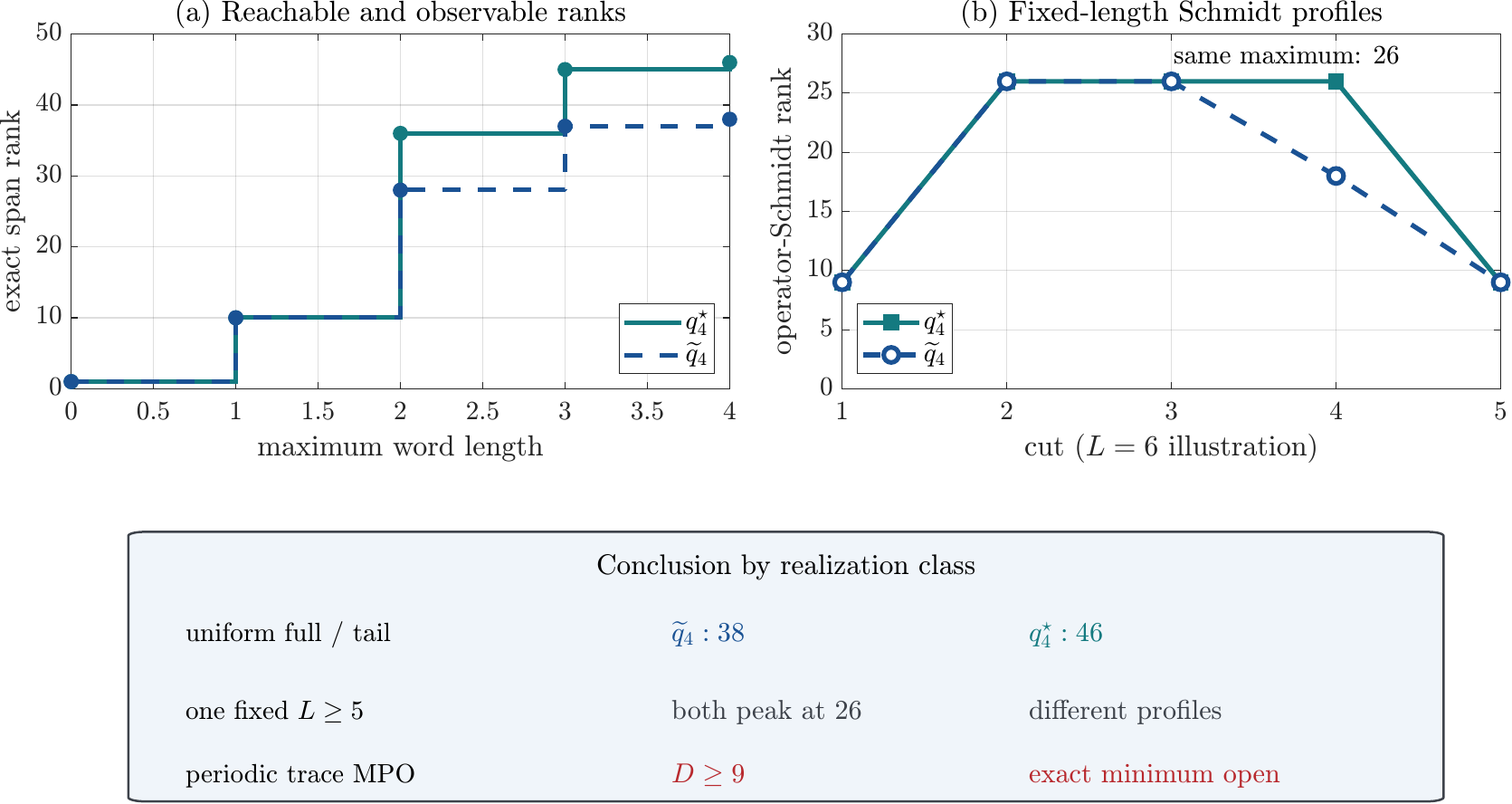}
\caption{Realization-dependent complexity. (a) Reachable and observable ranks saturate at \(38\) for \(\widetilde q_4\) and \(46\) for \(q_4^\star\); tail-completion rigidity gives the same minima when shorter sectors are free. (b) At \(L=6\), the fixed-length profiles differ while sharing maximum rank \(26\); line styles and markers provide color-independent identification. The conclusion box separates these uniform and fixed-length results from the periodic trace-MPO lower bound \(D\ge9\), whose exact minimum remains open.}
\label{fig:complexity-classes}
\end{figure}

Lemma~\ref{lem:fixed-all-lengths}, Proposition~\ref{prop:fixed-representative-comparison}, and Theorem~\ref{thm:full-series-minimum} solve different realization problems. Proposition~\ref{prop:periodic-trace-bound} supplies a lower bound for the periodic trace-MPO class. Blocked-site realizations and costs counted in SU(2) multiplets instead of ordinary auxiliary dimension are not compared here. Figure~\ref{fig:complexity-classes} compares the four realization problems for the reference densities.

\section{Discussion}\label{sec:discussion}

\subsection{What density gauge changes}

The transfer calculation fixes hierarchy-level normalization, whereas local alignment and Hilbert--Schmidt projections fix representative-level conventions. At fourth order the scalar offset of the boost density is removed by trace centering. At fifth order the transfer direction \(Q_{5,L}^{\mathrm{tr}}\) provides a convention-independent reference for the compact boost and orbit-aligned representatives. The width-four quotient theorem complements this normalization: it selects one periodic scalar hierarchy class after three-site divergences and lower directions have been removed, while leaving the choice of local representative open.

That remaining freedom is invisible to periodic summation but visible to a fixed-boundary tensor network. Reflection symmetrization maps \(\widetilde q_4\) to the divergence-equivalent \(q_4^\star\) without changing the periodic \(Q_4\), yet changes the local TT middle rank from \(18\) to \(26\) and the exact all-length realization dimension from \(38\) to \(46\). Tail rigidity shows that prescribed zeros at lengths below four do not cause this separation. The mechanism is instead a quotient-dimension obstruction in the long-word row space.

Optimizing the local gauge gives a stronger statement than this representative comparison. Across every Hermitian three-site divergence of the fixed four-site density, the full-series minimum is \(36\); imposing explicit local reflection symmetry raises it to \(44\). These bounds allow non-scalar densities, while the simple optimizers remain SU(2) scalars. The difference is the exact cost of a local reflection constraint in the stated representation class. The periodic charge and its physical symmetries are unchanged.

The comparison also clarifies which complexity is gauge sensitive. The complete fixed-length Schmidt profiles differ for every \(L\ge4\), although their maximum ranks agree at \(26\) for \(L\ge5\). The full and tail coefficient series retain the strict \(38\)-versus-\(46\) separation because one length-independent tensor and fixed boundaries must realize all required lengths simultaneously. The periodic trace-MPO problem has different boundary contractions and is constrained here only by the lower bound \(D\ge9\).

The fifth-order comparison provides a useful counterpoint. Alignment of the compact boost density and subsequent removal of its \(Q_3\) component change the exact five-site entries but leave all four TT ranks equal to \((9,33,33,9)\). Corollary~\ref{cor:q5-full-series} therefore gives the same full-series minimum \(86\) for each, with coefficients below length five set to zero. Representative changes need not alter local tensor ranks or this exact realization cost, and no monotonic relation between canonicalization and MPO compression should be expected. The fourth-order separation is a property of the gauge choice and realization class, not a generic consequence of changing density convention.

The affine-family classification illustrates a different feature. Along that line, dimension \(38\) occurs only at two reflection-related densities, while every other real parameter requires \(46\). Thus an arbitrarily small generic change along this gauge direction can increase the exact uniform realization cost, even though no point on the line attains the full gauge-class optimum. The truncated-bulk operators vary continuously with the parameter, but their exact minimal dimensions need not. This statement concerns exact representation; it does not imply a discontinuity in the cost of an approximation at fixed tolerance.

\subsection{Reusable MPOs, scope, and open problems}

The explicit densities provide inputs for exact diagonalization, tensor-network benchmarks, higher-current constructions, and integrable deformations such as \(H+\lambda_4Q_4+\lambda_5Q_5\). Open-chain applications require any necessary boundary counterterms to be supplied separately. The exact minima use ordinary auxiliary dimension for uniform unblocked one-site fixed-boundary MPOs. The full-series gauge optimization is complete within \(\mathcal G_4\) and \(\mathcal G_4^+\); optimization over larger density widths or over gauges in the tail-only problem remains open. A separate symmetry-adapted cost convention, blocked tensors, and the periodic trace-MPO minimum are different optimization problems.

The conserved-density classification is restricted to translation-invariant SU(2)-scalar densities of width at most four satisfying aligned five-site continuity. This restriction differs from the gauge optimization, which allows arbitrary Hermitian currents while fixing the periodic charge. Classifying non-scalar conserved charges, optimizing the five-site gauge class beyond the three fixed full-series realizations, and extending the explicit hierarchy beyond \(Q_5\) remain natural directions.

\section{Conclusions}\label{sec:conclusions}

Local-density gauge is invisible to a periodic conserved operator but remains visible to its boundary-sensitive MPO representation. For the spin-1 Takhtajan--Babujian fourth charge, optimization over all Hermitian three-site divergences of a fixed four-site density gives full-series minimum \(36\). Requiring the local density to be reflection even gives minimum \(44\). Explicit SU(2)-scalar densities with TT ranks \((8,17,9)\) and \((9,24,9)\) attain these bounds, and their individual tail minima agree. The proof combines a graded Hankel lemma with fixed coefficient-tensor invariants and a weighted-trace obstruction; it does not assume SU(2) symmetry of the competing densities.

The comparison is anchored in an explicit integrable hierarchy. We construct and transfer-normalize local representatives through \(Q_5\), including a reflection-even four-site density and a compact ten-word five-site boost density, and establish pairwise commutativity of the periodic \(Q_3,Q_4,Q_5\) on every admissible ring. In the translation-invariant SU(2)-scalar width-four sector, conserved densities modulo three-site divergences and the local \(I,H,Q_3\) directions form a one-dimensional reflection-even quotient generated by \(q_4^\star\). The transfer identification fixes the hierarchy direction, while the quotient theorem isolates its low-range scalar class.

The reference densities and their affine family illustrate why a representative minimum is not a gauge-class minimum: dimensions \(38\) and \(46\) on that line exceed the attainable \(36\), and the reflection-even reference density is also nonoptimal. The extensive periodic operator therefore does not determine the local data controlling a fixed-boundary realization. The results settle the stated four-site full-series problem, while larger support widths, gauge-global tail optimization, and periodic trace-MPO minima remain open.

The graded Hankel construction also fixes the full-series minimum at \(86\)
for each of the three displayed five-site densities, under the zero
short-length convention. Together, the explicit local identities, universal
four-site lower bounds, and finite Hankel witnesses separate hierarchy
identification from the cost of representing a chosen density. Their
finite algebraic inputs can be reconstructed individually from the
Supplementary Computational Material.
\appendix
\numberwithin{equation}{section}
\renewcommand{\theequation}{\Alph{section}.\arabic{equation}}
\numberwithin{table}{section}
\renewcommand{\thetable}{\Alph{section}.\arabic{table}}

\section{Spin-1 projectors and integrability conventions}\label{app:spin1}

\subsection{One-site and two-site operators}

In the ordered basis \((|+1\rangle,|0\rangle,|-1\rangle)\),

\[
\begin{aligned}
S^x&=\frac1{\sqrt2}
\begin{pmatrix}0&1&0\\1&0&1\\0&1&0\end{pmatrix},
&\qquad
S^y&=\frac1{\sqrt2}
\begin{pmatrix}0&-i&0\\i&0&-i\\0&i&0\end{pmatrix},\\
S^z&=\begin{pmatrix}1&0&0\\0&0&0\\0&0&-1\end{pmatrix}.
\end{aligned}
\]

For \(X=\boldsymbol S_1\cdot\boldsymbol S_2\), the total-spin sectors \(s=0,1,2\) have eigenvalues \(-2,-1,1\). Lagrange interpolation therefore gives

\[
\begin{aligned}
P_0&=\frac{(X+1)(X-1)}3,
&\qquad
P_1&=-\frac{(X+2)(X-1)}2,\\
P_2&=\frac{(X+2)(X+1)}6.
\end{aligned}
\]

They obey \(P_sP_t=\delta_{st}P_s\), \(\sum_sP_s=I\), and the two-site permutation is

\[
P=P_0-P_1+P_2=X^2+X-I.
\]

The local Hamiltonian is \(h=X-X^2\), with channel eigenvalues \((-6,-2,0)\).

\subsection{\texorpdfstring{Fused \(R\) matrix and Hamiltonian derivative}{Fused R matrix and Hamiltonian derivative}}

The polynomial fused matrix used throughout the calculation is

\[
R(u)=(u-1)(u-2)P_0+(u+1)(u-2)P_1+(u+1)(u+2)P_2.
\]

At the regular point,

\[
R(0)=2(P_0-P_1+P_2)=2P,
\]

while \(R'(0)\) has channel eigenvalues \((-3,-1,3)\). Hence

\[
2R^{-1}(0)R'(0)-3I
=(-6)P_0+(-2)P_1+0P_2=X-X^2=h.
\]

The polynomial Yang--Baxter identity is checked in the convention

\[
R_{12}(u)R_{13}(u+v)R_{23}(v)
=R_{23}(v)R_{13}(u+v)R_{12}(u).
\]

After clearing the \(1/\sqrt2\) factors in the spin matrices, every coefficient in \(u\) and \(v\) is an integer matrix. Exact expansion gives zero for every bivariate coefficient matrix; independent Mathematica and MATLAB calculations give the same result.

\subsection{Lowest local integrability identity}

Expanding the Yang--Baxter equation at the regular point, or multiplying the three-site matrices directly, yields

\[
[h_j+h_{j+1},[h_j,h_{j+1}]]=h_j^2-h_{j+1}^2.
\]

This is the Reshetikhin identity in the normalization of this paper. It fixes the local current in Sec.~\ref{sec:q3-continuity} without invoking a cut-dependent finite-ring boost. Transfer matrices are defined by

\[
t_L(u)=\operatorname{tr}_aR_{aL}(u)\cdots R_{a1}(u),
\]

and their commutativity supplies the formal conserved hierarchy. Equations~\eqref{eq:transfer-charge-definition}--\eqref{eq:q5-transfer-map} use centered logarithmic derivatives to fix the scalar convention at fourth and fifth order and give the exact relation to the displayed local representatives.

\section{\texorpdfstring{Detailed derivation of \(q_4^\star\)}{Detailed derivation of q4 star}}\label{app:q4}

\subsection{Current for the three-site charge}

Set \(C_j=[h_j,h_{j+1}]\). The Reshetikhin identity, the Jacobi identity, and \([h_j,h_{j+2}]=0\) give

\[
[H,C_j]=F_j-F_{j+1},\qquad
F_j=[h_{j-1},C_j]+h_j^2.
\]

Thus \(Q_3=i\sum_jC_j\) is conserved by periodic telescoping whenever the four sites in the local identity remain distinct.

\subsection{Removing the absolute boost coordinate}

In the infinite-chain local algebra take \(B=\sum_k kh_k\) and \(Q_4=-i[B,Q_3]\). Only \(k=j-1,j,j+1,j+2\) overlap a fixed \(C_j\), so

\[
\begin{aligned}
Q_4={}&\sum_jj[H,C_j]\\
&+\sum_j\left(-[h_{j-1},C_j]+[h_{j+1},C_j]
+2[h_{j+2},C_j]\right).
\end{aligned}
\]

In the formal infinite-chain algebra, the first moment of the divergence satisfies

\[
\sum_jj(F_j-F_{j+1})=\sum_jF_j.
\]

Its \([h_{j-1},C_j]\) contribution cancels the first relative-coordinate term. Therefore

\[
Q_4=\sum_jg_j,
\]

with

\[
g_j=h_j^2+[h_{j+1},[h_j,h_{j+1}]]
+2[h_{j+2},[h_j,h_{j+1}]].
\]

This is a formal density representative of the boost descendant. With the phase convention of Proposition~\ref{prop:hierarchy-normalization}, the corresponding raw Hokkyo representative is \(-\sum_jg_j\). Its periodic embedding has
\[
\frac{\operatorname{Tr}Q_L^{\mathrm{per}}[g]}{3^L}
=\frac{16}{3}L,
\]
so commutator recursion alone does not select the identity component needed for comparison with a centered transfer charge. The transfer convention fixes it as
\[
\widehat Q_L^{(4),\mathrm{per}}
=-Q_L^{\mathrm{per}}[g]+\frac{16}{3}LI_L
=8\overline T_{4,L}.
\]
This distinction is also why applying \(\sum_jjh_j\) directly on a finite periodic ring is not the definition used here.

\subsection{Four-site reflection and Hilbert--Schmidt fixing}

On four sites write \(a=h_{12}\), \(b=h_{23}\), \(c=h_{34}\), so that

\[
g=a^2+[b,[a,b]]+2[c,[a,b]].
\]

Exact five-site embedding gives

\[
\sum_jg_j=\sum_j(\mathcal Rg\mathcal R)_j.
\]

Consequently

\[
e=\frac{g+\mathcal Rg\mathcal R}{2}
\]

is a reflection-even density of the same periodic operator. Define \(d_H=(a+c)/2\), whose bulk translation sum is the Hamiltonian up to the usual boundary convention. The exact four-site Hilbert--Schmidt data are

\[
\begin{array}{c|cc}
 & I & d_H\\ \hline
e   & 432  & -2160\\
d_H & -108 & 288
\end{array},\qquad \operatorname{Tr}I=81.
\]

Solving \(\langle e+\alpha d_H+\beta_4 I,I\rangle_{\mathrm{HS}}=0\) and \(\langle e+\alpha d_H+\beta_4 I,d_H\rangle_{\mathrm{HS}}=0\) gives

\[
\alpha=11,\qquad \beta_4=\frac{28}{3}.
\]

With the chosen overall scale,

\[
q_4^\star=-\frac1{36}\left[e+11d_H+\frac{28}{3}I\right],
\]

or equivalently

\[
216q_4^\star=-\left[3(g+\mathcal Rg\mathcal R)+33(a+c)+56I\right].
\]

It follows exactly that

\[
\operatorname{Tr}q_4^\star=0,\qquad
\|q_4^\star\|_{\mathrm{HS}}^2=39.
\]

Define

\[
Q_L^{\mathrm{per}}[g]:=\sum_{j=1}^{L}g_j,\qquad
Q_{4,L}^{\star,\mathrm{per}}
:=Q_L^{\mathrm{per}}[q_4^\star].
\]

The corresponding periodic operators satisfy

\begin{equation}\label{eq:q4-periodic-relation-app}
Q_L^{\mathrm{per}}[g]
+11H_L+\frac{28}{3}L I_L
+36Q_{4,L}^{\star,\mathrm{per}}=0.
\end{equation}
\subsection{Exact finite-ring completion}

Exact integer arithmetic verifies Eq.~\eqref{eq:q4-continuity}. The six-site local-divergence argument is unaliased for \(L\ge6\). At \(L=4,5\), independent integer-matrix calculations give

\begin{equation}\label{eq:q4-small-ring-completion}
[H_L,Q_{4,L}^{\star,\mathrm{per}}]=0,
\qquad L=4,5,
\end{equation}
and both commutators vanish exactly. Independent calculations also confirm the polynomial Yang--Baxter equation, Hamiltonian derivative, Reshetikhin identity, Hermiticity, reflection parity, SU(2) invariance, tracelessness, and the boost/lower-charge relation.

\section{Five-site recursion, alignment, and hierarchy normalization}\label{app:q5}

\subsection{Product basis and the explicit alignment coboundary}

With \(E_{ab}:=|a\rangle\langle b|\), the real one-site basis used for orbit alignment is
\[
\begin{aligned}
B_0&=I,\\
B_1&=\operatorname{diag}(1,-1,0),&
B_2&=\operatorname{diag}(1,1,-2),\\
B_3&=E_{01}+E_{10},&
B_4&=E_{01}-E_{10},\\
B_5&=E_{02}+E_{20},&
B_6&=E_{02}-E_{20},\\
B_7&=E_{12}+E_{21},&
B_8&=E_{12}-E_{21}.
\end{aligned}
\]
Its squared Hilbert--Schmidt norms are \((3,2,6,2,2,2,2,2,2)\). If \(Y_\gamma\) is the left-aligned active core of a finite-support translation orbit, write that orbit's contribution as
\[
A_\gamma=\sum_{s=0}^{r-|\gamma|}c_{\gamma,s}T^sY_\gamma.
\]
Left alignment replaces this expression by \((\sum_s c_{\gamma,s})Y_\gamma\), and therefore
\begin{equation}\label{eq:left-alignment-coboundary}
\mathcal L_r(A)-A=(1-T)D_A,
\qquad
D_A=\sum_{\gamma,s}c_{\gamma,s}
\sum_{t=0}^{s-1}T^tY_\gamma .
\end{equation}
The inner sum vanishes for \(s=0\). With reflection translated back to the same origin, \(\mathcal RTR=T^{-1}\), so
\[
\mathcal R(1-T)D_A\mathcal R
=-(1-T)T^{-1}\mathcal R D_A\mathcal R.
\]
For reflection-odd \(A\), Eq.~\eqref{eq:orbit-alignment} consequently gives Eq.~\eqref{eq:alignment-divergence} with
\[
D_A^{(-)}
=\frac12\left(D_A+T^{-1}\mathcal R D_A\mathcal R\right).
\]
This proves the finite-support identity used in Lemma~\ref{lem:alignment-preserves-charge}, before any periodic summation is taken.

\subsection{Primitive lower-charge-fixed numerator}

For a five-site rational coefficient tensor \(A\), define

\[
N[A]=2\cdot6^5\,
\operatorname{Align}_{\mathrm{orb},5}^{(-)}(A)
=15552\,
\operatorname{Align}_{\mathrm{orb},5}^{(-)}(A).
\]

Let

\[
N_5=N[A_{72}],
\qquad
N_3=N[A_3^{(5)}].
\]

The integer-numerator projection coefficient is

\[
\frac{\langle N_5,N_3\rangle_{\mathrm{HS}}}
{\langle N_3,N_3\rangle_{\mathrm{HS}}}
=\frac{1102}{11},
\]

which becomes \(551/396\) after the \(1/72\) Hermitian normalization. The orthogonalized numerator

\[
M=11N_5-1102N_3
\]

has entrywise greatest common divisor \(93312\), giving

\begin{equation}\label{eq:q5-integer-app}
A_5^\star
=\frac{11N_5-1102N_3}{93312},
\qquad
q_5^\star=\frac{i}{132}A_5^\star.
\end{equation}
The map used here is

\[
\operatorname{Align}_{\mathrm{orb},5}^{(-)}
=\frac{I-\mathcal R}{2}\,\mathcal L,
\]

where \(\mathcal L\) left-aligns every translation orbit. Its exact action on the \(9^5=59049\)-dimensional rational coefficient space gives the diagnostics in Table~\ref{tab:alignment-diagnostics}.

\begin{table}[tbp]
\centering
\small
\caption{Exact diagnostics for the translation-orbit alignment map and the three outputs used in the construction of \(q_5^\star\).}
\label{tab:alignment-diagnostics}
\begin{tabularx}{\textwidth}{@{}>{\raggedright\arraybackslash}X>{\raggedright\arraybackslash}p{0.30\textwidth}@{}}
\toprule
diagnostic & exact result \\
\midrule
\(\mathcal L^2=\mathcal L\) & yes \\
\(\mathcal R^2=I\) & yes \\
aligned \(Q_5\) output is reflection odd & yes \\
aligned \(Q_5\) output is a fixed point & yes \\
embedded \(Q_3\) output is a fixed point & yes \\
\(A_5^\star\) is a fixed point & yes \\
\bottomrule
\end{tabularx}
\end{table}

Only the action on the displayed outputs is used below. We therefore treat \(\operatorname{Align}_{\mathrm{orb},5}^{(-)}\) as an alignment convention and do not require it to be a projector on the complete coefficient space. All three outputs used here are exact fixed points.

\subsection{From the conditional recursion to the compact density}\label{app:q5-recursion-details}

Here all matrix indices refer to one six-site window, ordered from left to right. The density \(G_{216}=216q_4^\star\) occupies sites \(2,3,4,5\); its current occupies sites \(1,\ldots,5\). This fixes the origin in the relative-coordinate weights of Eq.~\eqref{eq:conditional-boost-density}. Define
\[
F_6=\sum_{s=-1}^{3}[h_{s+2,s+3},I\otimes G_{216}\otimes I],
\qquad
W_6=\sum_{s=-1}^{3}s[h_{s+2,s+3},I\otimes G_{216}\otimes I].
\]
Both are real antisymmetric integer matrices. The Hamiltonian is constructed from the one-site spin matrices, and \(G_{216}\) is reconstructed from Eq.~\eqref{eq:q4-density}.

To obtain the current without a numerical fit, use the real orthogonal basis \(B_\mu\) defined above. Let \(n_\mu=\operatorname{Tr}(B_\mu^{\mathsf T}B_\mu)\), so \((n_0,\ldots,n_8)=(3,2,6,2,2,2,2,2,2)\). For a six-site integer matrix \(X\), the scaled coefficients
\begin{equation}\label{eq:recursion-product-coefficients}
c_{\boldsymbol\mu}(X)=
6^6\frac{\operatorname{Tr}[(B_{\mu_1}\otimes\cdots\otimes B_{\mu_6})^{\mathsf T}X]}
{n_{\mu_1}\cdots n_{\mu_6}}
\end{equation}
are integers. In each orbit of a nonempty active core \(\gamma\) of length \(m\), list its placement coefficients as \(c_0,\ldots,c_{6-m}\). A divergence requires their sum to vanish. If it does, the cumulative coefficients
\[
j_t=\sum_{s=0}^{t}c_s,\qquad 0\le t\le5-m,
\]
give a five-site matrix \(\mathcal J[X]\) satisfying
\[
6^6X=\mathcal J[X]\otimes I-I\otimes\mathcal J[X].
\]
The empty-core coefficient of the current is set to zero, fixing the additive identity freedom. This is the traceless current convention. Zero orbit sums and the reconstructed matrix equality are checked separately.

For \(X=F_6\), exact integration gives
\[
\mathcal J[F_6]=3\,6^6K_4^{[72]}.
\]
The division is entrywise exact. Thus \(F_6=3\Delta K_4^{[72]}\), which is Eq.~\eqref{eq:q4-continuity}, and the physical current is \(J^{(4)}=K_4^{[72]}/72\). Substituting this current into Eq.~\eqref{eq:conditional-boost-density} yields
\[
q_{5,\mathrm{rec}}^{(6)}
=-i\left(\frac{K_4^{[72]}\otimes I}{72}+\frac{W_6}{216}\right)
=\frac{i}{72}\left(-K_4^{[72]}\otimes I-\frac{W_6}{3}\right).
\]
The weighted commutator \(W_6\) is also entrywise divisible by three. This establishes the explicit six-site numerator of Eq.~\eqref{eq:q5-recursion-explicit} from the recursion inputs.

Independently multiply the bond operators in the ten words of Eq.~\eqref{eq:q5-word} to construct \(A_{72}\). On the same six-site window it occupies sites \(1,\ldots,5\). Apply the orbit integration once more to
\[
X=\widehat A_{72}^{(6)}-A_{72}\otimes I.
\]
All orbit sums vanish, and \(K_{\mathrm B}=\mathcal J[X]\) gives Eq.~\eqref{eq:q5-boost-identity}. This calculation establishes the local identity before any periodic translation sum is taken. The reconstructed \(K_4^{[72]}\), \(\widehat A_{72}^{(6)}\), \(A_{72}\), and \(K_{\mathrm B}\) are real antisymmetric, with dimensions \(243\), \(729\), \(243\), and \(243\), respectively.

Exact integer arithmetic verifies the four-site continuity equation and the six-site recursion identity entry by entry, with zero residual. The reconstructed matrices also agree exactly with the archived matrices. These are comparison targets; neither an archived recursion output nor an archived difference is used to construct either side of the identity. The six-site identity supplies the periodic equivalence for all \(L\ge6\) by telescoping. The \(L=5\) checks concern the compact operator itself.

\subsection{Transfer-normalization details}\label{app:transfer-normalization-details}

The scalar shift used in Proposition~\ref{prop:hierarchy-normalization} gives \(\operatorname{Tr}_1h^{(0)}=\operatorname{Tr}_2h^{(0)}=0\), because SU(2) invariance yields \(\operatorname{Tr}_1h=\operatorname{Tr}_2h=-4I_3\). It changes neither commutators nor the non-scalar boost descendants from third order onward. Comparing recursions in the formal local algebra gives \(Q_n=(-i)^{n-2}\widehat Q^{(n)}\) before representative and lower-charge changes, up to an identity component not fixed by commutators. The third charge is traceless, giving Eq.~\eqref{eq:hierarchy-map-q3}.

For the polynomial \(R\) matrix,
\begin{equation}\label{eq:R-integer-polynomial}
R(u)=2P+uP(h+3I)+u^2I,
\qquad
R(0)^{-1}R'(0)=\frac12(h+3I)
=\frac12h^{(0)}+\frac56I_9.
\end{equation}
The scalar term drops out of boost commutators, hence \(B_{\mathrm{tr}}=B_{H^{(0)}}/2\). Regularity, the difference-form Yang--Baxter equation, and the nearest-neighbor logarithmic derivative allow the standard ladder theorem \cite{R07,R08} to be applied. After normalized-trace subtraction,
\[
\overline T_{n+1}=[B_{\mathrm{tr}},\overline T_n].
\]
Direct differentiation gives \(2\overline T_2=H^{(0)}\), and induction yields
\begin{equation}\label{eq:boost-transfer-general-map}
\widehat Q^{(n)}=2^{n-1}\overline T_n.
\end{equation}

The raw fourth-order boost density has periodic sum \(-Q_L^{\mathrm{per}}[g]\), and \(\operatorname{Tr}g=432\) gives
\[
\frac{\operatorname{Tr}Q_L^{\mathrm{per}}[g]}{3^L}
=\frac{16}{3}L.
\]
Removing this scalar proves Eq.~\eqref{eq:q4-transfer-map}. Periodic summation of Eq.~\eqref{eq:q4-density} further gives
\[
36Q_{4,L}^{\star,\mathrm{per}}
=-Q_L^{\mathrm{per}}[g]-11H_L-\frac{28}{3}LI_L
=\widehat Q_L^{(4),\mathrm{per}}-11H_L^{(0)}.
\]
Applying one more formal boost step and using Eq.~\eqref{eq:q5-boost-identity} gives Eq.~\eqref{eq:hierarchy-map-q5boost} on unaliased embeddings \(L\ge6\). Equation~\eqref{eq:q5-hierarchy} and \(-11/36-551/396=-56/33\) then give Eq.~\eqref{eq:hierarchy-map-q5star}.

Independent polynomial transfer-matrix expansions verify the scalar and ladder normalization:
\begin{equation}\label{eq:q4-transfer-finite-check}
8\overline T_{4,L}
=-Q_L^{\mathrm{per}}[g]+\frac{16}{3}LI_L,
\qquad L=4,5.
\end{equation}
Writing \(A_3=[h_{12},h_{23}]\), \(A_{72,L}=Q_L^{\mathrm{per}}[A_{72}]\), and \(A_{3,L}=Q_L^{\mathrm{per}}[A_3]\), the fifth-order check is
\begin{equation}\label{eq:q5-transfer-finite-check}
T_{5,L}
=-\frac1{32}A_{72,L}
-\frac{11}{16}A_{3,L}
-\frac{51L}{8}I_L,
\qquad L=6,7,
\end{equation}
as an exact matrix identity.

\subsection{Small-ring nontriviality data}\label{app:q5-small-ring-data}

For \(L=5,6\), exact integer arithmetic gives

\[
\begin{array}{c|ccc}
L&
\|Q_{3,L}^{\mathrm{per}}\|_{\mathrm{HS}}^2&
\langle Q_{3,L}^{\mathrm{per}},Q_{5,L}^{\star,\mathrm{per}}\rangle_{\mathrm{HS}}&
\|Q_{5,L}^{\star,\mathrm{per}}\|_{\mathrm{HS}}^2\\ \hline
5&23760&0&793740/11\\
6&85536&0&2363256/11
\end{array}.
\]

The displayed two-charge Gram matrices have rank two. Exact Gram elimination also gives rank four for each ordered set \(\{H,I,Q_3,Q_4^\star\}\) and \(\{H,Q_3,Q_4^\star,Q_5^\star\}\) at both \(L=5\) and \(L=6\), as used in Corollary~\ref{cor:hierarchy-independence}.

\section{Product-word orbit integration}\label{app:orbit-integration}

Let \(\jmath_p^{(n)}(A)\) embed an \(m\)-site operator \(A\) in an \(n\)-site window with \(p\) identities to its left. Place \(A_3=[h_{12},h_{23}]\) at relative origin zero and a five-site numerator \(A_5\) at relative origin \(s\in\{-4,\ldots,2\}\). With \(o_s=\min(0,s)\), define
\begin{equation}\label{eq:M7}
\mathcal M_7[A_3,A_5]
:=\sum_{s=-4}^{2}
\left[
\jmath_{-o_s}^{(7)}(A_3),
\jmath_{s-o_s}^{(7)}(A_5)
\right].
\end{equation}
It is aligned so that
\[
\sum_j\tau_j\!\left(\mathcal M_7[A_3,A_5]\right)
=\left[\sum_jA_{3,j},\sum_kA_{5,k}\right].
\]

For a seven-site coefficient tensor, group all placements of an active core \(\gamma\). If their integer coefficients are

\[
c_0,c_1,\ldots,c_{7-m},
\]

a necessary and sufficient condition for a one-site divergence within that orbit is

\[
\sum_{s=0}^{7-m}c_s=0.
\]

When this holds, define cumulative current coefficients

\[
j_t=\sum_{s=0}^{t}c_s,
\qquad t=0,\ldots,6-m.
\]

Then

\[
c_0=j_0,\qquad
c_s=j_s-j_{s-1},\qquad
c_{7-m}=-j_{6-m},
\]

which reconstructs

\begin{equation}\label{eq:orbit-divergence-app}
\mathcal D_7=J_6\otimes I-I\otimes J_6.
\end{equation}
The exact orbit integration gives the three identities summarized in Table~\ref{tab:local-divergence-diagnostics}. All coefficients are evaluated after the common scale \(6^7=279936\).

\begin{table}[htbp]
\centering
\small
\caption{Exact local-divergence identities after the common scale \(6^7=279936\).}
\label{tab:local-divergence-diagnostics}
\begin{tabular}{@{}lcc@{}}
\toprule
local identity & current factor & exact divergence reconstruction \\
\midrule
\([H,A_{72}]\) & \(1/3\) & verified \\
\([H,A_5^\star]\) & \(1/3\) & verified \\
\([Q_3,A_5^\star]\) & \(1\) & verified \\
\bottomrule
\end{tabular}
\end{table}

For the first two rows, after reducing the common integer factor,

\[
\mathcal D_7[A]
=\frac13(\mathcal K_6\otimes I-I\otimes \mathcal K_6).
\]

For the mutual commutator the denominator is one. Hence the Hermitian normalizations in the main text give

\[
[H,q_{5,j}^{\mathrm{boost}}]
=\frac{i}{216}(\mathcal K_{6,j}^{\mathrm{boost}}\otimes I
-I\otimes \mathcal K_{6,j}^{\mathrm{boost}}),
\]

\[
[H,q_{5,j}^{\star}]
=\frac{i}{396}(\mathcal K_{6,j}^{\star}\otimes I
-I\otimes \mathcal K_{6,j}^{\star}).
\]

For \(L\ge7\) these identities telescope without support aliasing. The two remaining lengths in the theorem are covered by the independent exact integer checks in Table~\ref{tab:q5-small-ring-checks}:

\begin{table}[tbp]
\centering
\small
\caption{Exact small-ring commutator checks completing the local \(Q_5\) continuity arguments.}
\label{tab:q5-small-ring-checks}
\begin{tabularx}{\textwidth}{@{}>{\raggedright\arraybackslash}p{0.18\textwidth}*{2}{>{\raggedright\arraybackslash}X}@{}}
\toprule
commutator identity & \(L=5\) & \(L=6\) \\
\midrule
\([H_L,Q_{5,L}^{\mathrm{boost}}]\) & 0 & 0 \\
\([H_L,Q_{5,L}^{\star}]\) & 0 & 0 \\
\([Q_{3,L},Q_{5,L}^{\star}]\) & 0 & 0 \\
\bottomrule
\end{tabularx}
\end{table}

These matrices are reconstructed from the one-site spin operators and integer inputs. Exact sparse arithmetic also gives \([Q_3,Q_4]=0\) for \(L=4,\ldots,9\) and \([Q_4,Q_5^{\mathrm{boost}}]=[Q_4,Q_5^\star]=0\) for \(L=5,\ldots,9\). Corollary~2 of Ref.~\cite{R42}, with integer levels \(2\le n,m\le\lfloor N/2\rfloor\), supplies \([Q_3,Q_4]=0\) for \(L\ge8\) and \([Q_4,Q_5]=0\) for \(L\ge10\), completing the large-ring ranges.

\section{Exact MPO realization and minimality}\label{app:mpo}

\subsection{Exact TT cores and the uniform tensor}

Expand the integer-normalized density in matrix units,

\[
216q_4^\star
=\sum_{\mu_1,\ldots,\mu_4}
G_1^{\mu_1}G_2^{\mu_2}G_3^{\mu_3}G_4^{\mu_4}
E_{\mu_1}\otimes\cdots\otimes E_{\mu_4}.
\]

Exact rational flattening gives core shapes

\[
G_1^\mu:1\times9,\quad
G_2^\mu:9\times26,\quad
G_3^\mu:26\times9,\quad
G_4^\mu:9\times1.
\]

Let \(\iota_\mu=1\) for \(\mu\in\{0,4,8\}\) and zero otherwise, so that \(\sum_\mu\iota_\mu E_\mu=I\). On the virtual direct sum

\[
\mathbb Q^{46}
=\mathbb Q_{\mathrm{idle}}
\oplus\mathbb Q^9\oplus\mathbb Q^{26}\oplus\mathbb Q^9
\oplus\mathbb Q_{\mathrm{done}},
\]

define

\begin{equation}\label{eq:uniform-mpo-app}
W^\mu=
\begin{pmatrix}
\iota_\mu&G_1^\mu&0&0&0\\
0&0&G_2^\mu&0&0\\
0&0&0&G_3^\mu&0\\
0&0&0&0&G_4^\mu\\
0&0&0&0&\iota_\mu
\end{pmatrix}.
\end{equation}
With

\[
\ell=\frac1{216}(1,0,\ldots,0),\qquad
r=(0,\ldots,0,1)^\mathsf T,
\]

each accepting path inserts exactly one four-site density and propagates the identity before and after it. The nine \(46\times46\) matrices contain 598 nonzero rational entries in sparse triplet form; the machine-readable tensors include the boundaries, physical indices, numerators, and denominators.

No path can travel from idle to done in fewer than four physical letters. Thus the same tensor gives zero coefficients exactly at \(L=0,1,2,3\), not merely after operator cancellation.

\subsection{Reachability, observability, and the Hankel witness}

Starting with the empty word and adjoining words up to length four, the cumulative exact reachable-space dimensions are

\[
(1,10,36,45,46).
\]

The observable-space dimensions obtained from the right boundary are the same. A greedily independent set of 46 prefix and 46 suffix words, all of lengths between zero and four, gives the finite Hankel witness.

Let

\[
\mathsf R_{i,:}=\ell W^{u_i},\qquad \mathsf O_{:,j}=W^{v_j}r.
\]

Exact rational elimination gives

\[
\operatorname{rank}\mathsf R=\operatorname{rank}\mathsf O=46.
\]

The corresponding finite Hankel matrix is

\begin{equation}\label{eq:hankel-factor-app}
\mathsf H_{ij}=f_4(u_iv_j)=(\mathsf R\mathsf O)_{ij},
\end{equation}
and has exact rank 46. Since every \(D\)-dimensional realization factors every finite Hankel submatrix through a \(D\)-dimensional space, \(D\ge46\). The explicit tensor above attains the bound, proving Theorem~\ref{thm:full-series-minimum} over \(\mathbb Q\). The same nonzero rational minor remains nonzero over \(\mathbb R\) and \(\mathbb C\).

Repeating the exact TT and finite-state construction for \(\widetilde q_4\) gives ranks \((9,18,9)\), dimension \(38\), and cumulative reachable/observable dimensions \((1,10,28,37,38)\). A rank-38 Hankel witness proves minimality. Exact reconstruction of Eq.~\eqref{eq:q4-representative-divergence} confirms that the 38-versus-46 comparison concerns two representatives of the same periodic charge.

\subsection{Fixed length and tail realization}

Exact-length row and column spaces have dimensions

\[
(1,9,26,35,36,36,\ldots).
\]
Their exact contractions give
\[
\operatorname{SchmidtProfile}(\mathcal B_{4,L}[q_4^\star])
=\left(9,\underbrace{26,\ldots,26}_{L-3},9\right)
\quad(L\ge4).
\]
Therefore every site-dependent fixed-boundary MPO for a fixed \(L\ge4\) has minimum possible maximum bond dimension 26. This does not contradict either uniform full-series value.

For a tail competitor, the fixed length-four rows span a 36-dimensional space \(S\), while their restriction to columns of length at least four has dimension 26. The fixed restricted rows of lengths at most four span dimensions 36 and 28 for \(q_4^\star\) and \(\widetilde q_4\), respectively. If the complete competitor row space has dimension \(R\), the quotient argument of Theorem~\ref{thm:tail-rigidity} gives \(R\ge36+36-26=46\) or \(R\ge36+28-26=38\). Both bounds are attained by the explicit rational realizations.

For the separate periodic trace-MPO problem, exact reshaping gives \(\operatorname{SchmidtProfile}(Q_{4,4}^{\star,\mathrm{per}})=(9,74,9)\). Since a trace-MPO of dimension \(D\) has operator-Schmidt rank at most \(D^2\), this proves \(D\ge9\); its exact minimum remains open. Costs based on blocked tensors or on SU(2) multiplet counts are not analyzed here.

\subsection{Polynomial Hankel witnesses for the affine family}\label{app:affine-hankel}

Use the integer density \(216q_4(\lambda)\) and denote its full-series coefficient function by \(f_\lambda\). For either fixed prefix/suffix list, every Hankel entry is affine:
\[
H_{ij}(\lambda)=f_\lambda(u_iv_j)
=(1-\lambda)f_{\widetilde q}(u_iv_j)+\lambda f_{q^\star}(u_iv_j),
\]
where both endpoint functions use the scale \(216\). The two lists supplied in the computational material each have 46 prefixes and 46 suffixes. Their matrices are obtained both from the MPO tensors and directly from density coefficients. In the latter route, for a word \(w\) of length \(L\),
\[
f_\lambda(w)=\sum_{j=0}^{L-4}
c_\lambda(w_{j+1}\cdots w_{j+4})
\prod_{k\notin\{j+1,\ldots,j+4\}}\iota_{w_k},
\]
with an empty sum for \(L<4\). Here \(c_\lambda\) is the four-letter coefficient tensor and \(\iota\) is the identity-symbol indicator used in Eq.~\eqref{eq:uniform-mpo-app}.

Normalize each nonzero determinant polynomial to be monic. Exact polynomial elimination gives
\begin{equation}\label{eq:family-Hankel-polynomials}
\begin{aligned}
p_A(\lambda)&=\frac{\lambda^9(\lambda-2)^9(4\lambda-11)(4\lambda+3)}{16},\\
p_B(\lambda)&=\frac{\lambda^8(\lambda-2)^8}{451584}
(24\lambda-37)(24\lambda-11)\\
&\hspace{2em}\times(28\lambda-45)(28\lambda-11).
\end{aligned}
\end{equation}
Their monic greatest common divisor is
\[
g(\lambda)=\lambda^8(\lambda-2)^8.
\]
The remaining factors have disjoint roots; an exact B\'ezout identity for \(p_A/g\) and \(p_B/g\) is also supplied. Thus there is no untested algebraic parameter at which both witnesses lose rank, apart from \(0,2\). The full Hankel rank is at least 46 everywhere else.

For the upper bound, fraction-free elimination over \(\mathbb Q[\lambda]\) gives generic unfolding ranks \((9,26,9)\). Vanishing of larger minors is a polynomial identity and remains valid at every specialization, including values where an elimination pivot vanishes. The insertion construction therefore supplies \(D\le46\) uniformly in the parameter. At \(0,2\), exact TT ranks and separate rank-38 Hankel witnesses give equality at 38. These arguments prove Theorem~\ref{thm:affine-gauge-family}; no numerical tolerance or parameter-grid extrapolation is used.

The gauge parameter is real; it is not a Hamiltonian coupling. The polynomial rank argument also holds over \(\mathbb C\).

\subsection{Five-site full-series minima}\label{app:q5-mpo-upper-bounds}

The three \(243\times243\) exact matrices for \(q_5^{\mathrm{boost}}\), \(q_5^A\), and \(q_5^\star\) were expanded in the same nine-letter matrix-unit alphabet and flattened after sites \(1,2,3,4\). Exact ranks are
\[
\begin{array}{c|c|c}
\text{density} & \text{successive TT ranks} & D_{\min}^{\mathrm{full}}\\
\hline
q_5^{\mathrm{boost}}&(9,33,33,9)&86\\
q_5^{A}&(9,33,33,9)&86\\
q_5^\star&(9,33,33,9)&86
\end{array}.
\]
Exact elimination over \(\mathbb Q\) gives these ranks. An independent implementation performs sparse elimination over each of \(\operatorname{GF}(1009)\), \(\operatorname{GF}(1013)\), and \(\operatorname{GF}(1019)\). All four routes return the four ranks shown for every representative. These local ranks give the insertion upper bounds; the matching full-series lower bounds follow from Lemma~\ref{lem:graded-full-series}, as stated in Corollary~\ref{cor:q5-full-series}.

For each density, construct a finite witness as follows. Rebuild the boost
numerator from the spin Hamiltonian, then perform orbit alignment and the
stated lower-charge subtraction. The archived density matrices serve only as
comparison targets. In the matrix-unit alphabet, choose a nonsingular local minor
at each cut \(j=0,\ldots,5\). The prefix and suffix lengths are \(j\)
and \(5-j\), respectively. On collecting the six groups, blocks above
the diagonal have total length below five and vanish. Diagonal blocks
have length five and equal the selected local minors. The product of
their nonzero determinants is consequently the determinant of the
full \(86\times86\) witness.

Every Hankel entry is computed directly from the translated local density,
independently of an MPO contraction. The finite witness matrices and ordered
prefix/suffix lists are supplied for independent determinant checks.
The witnesses use integer numerators; multiplication by the nonzero
physical normalization, including \(i\), preserves the rank.
These are three fixed-density full-series minima, with coefficients
below length five prescribed to vanish.

\subsection{Exact-arithmetic methods}\label{app:crosswalk}

All finite-matrix identities and ranks use exact integer or rational arithmetic. Orbit integration reconstructs the local currents, sparse finite-ring calculations cover aliased short rings and all hierarchy commutators through \(L=9\), and polynomial transfer expansions verify Eqs.~\eqref{eq:q4-transfer-map} and \eqref{eq:q5-transfer-map} on \(L=4,5\) and \(L=6,7\), respectively. The width-four classification is rebuilt from the SU(2)-scalar bases, with its constraint rank independently reproduced modulo \(1009,1013,1019\).

The six-site recursion is reconstructed from the spin Hamiltonian and the
four-site density, while the compact five-site expression is constructed
separately from its bond words. For the four-site gauge bounds, a second
spin-matrix implementation independently verifies the coefficient tensors
and weighted-trace identity. Archived matrices serve as comparison targets,
not as construction inputs. The three fifth-order Hankel witnesses are computed
directly from the local densities, independently of MPO contraction.

The Supplementary Computational Material provides the finite densities,
currents, coefficient tensors, MPOs, and Hankel witnesses, together with
instructions for reproducing the calculations. The analytic arguments in the
text turn the finite identities into the all-length identification and the
gauge-class lower bounds.

\section{Proof of the four-site gauge-class minima}\label{app:gauge-optima}

\subsection{Hermitian coefficients and gauge invariants}

We use \(G_0=216q_4^\star\) from Eq.~\eqref{eq:q4-density}, with the fixed
Hamiltonian normalization of Eq.~\eqref{eq:local-h}. Write
\(e_{ab}=|a\rangle\langle b|\), with \(a,b=0,1,2\) in the physical order
\(|+1\rangle,|0\rangle,|-1\rangle\). An orthogonal Hermitian one-site basis is
\begin{align*}
H_0&=I,\\
H_1&=e_{01}+e_{10}+e_{12}+e_{21},&
H_2&=i(e_{01}-e_{10}+e_{12}-e_{21}),\\
H_3&=e_{00}-e_{22},&H_4&=e_{02}+e_{20},\\
H_5&=i(e_{02}-e_{20}),&H_6&=e_{01}+e_{10}-e_{12}-e_{21},\\
H_7&=i(e_{01}-e_{10}-e_{12}+e_{21}),&H_8&=e_{00}-2e_{11}+e_{22}.
\end{align*}
The squared norms \(n_a=\operatorname{Tr}H_a^2\) are
\((3,4,4,2,2,2,4,4,6)\). For every Hermitian \(G\in\mathcal G_4\),
\begin{equation}\label{eq:gauge-real-coefficients}
g_{abcd}=\frac{\operatorname{Tr}[(H_a\otimes H_b\otimes H_c\otimes H_d)G]}
{n_an_bn_cn_d}\in\mathbb R.
\end{equation}
Unless indicated otherwise, indices below run from \(1\) to \(8\), excluding
the identity. Under \(G\mapsto G+\Delta_3C\), the coefficient change is
\(\delta g_{abcd}=c_{abc}\delta_{d0}-\delta_{a0}c_{bcd}\).
Consequently the following tensors are gauge invariant:
\begin{equation}\label{eq:gauge-invariant-tensors}
\begin{aligned}
A_{abcd}&=g_{abcd},& B_{abc}&=g_{abc0}+g_{0abc},\\
M_{ab}&=g_{a0b0}+g_{0a0b},&
N_{ab}&=g_{ab00}+g_{0ab0}+g_{00ab}.
\end{aligned}
\end{equation}
They describe four active sites, three consecutive active sites, separated
pairs, and adjacent pairs. All coefficients with active first and last indices
and an interior identity vanish. The one-active-site orbit sums and the
all-identity coefficient also vanish. Expansion of \(G_0\) establishes these
facts; the coefficient change above shows that they hold throughout
\(\mathcal G_4\).

The pair matrices and a positive diagonal weight are
\begin{equation}\label{eq:gauge-pair-matrices}
\begin{aligned}
M&=\operatorname{diag}(-21,-21,-42,-18,-18,-9,-9,-6),\\
N&=\operatorname{diag}(-15/2,-15/2,-15,-27,-27,-27/2,-27/2,-9),\\
\Omega&=\operatorname{diag}(66,66,33,17,17,34,34,51).
\end{aligned}
\end{equation}
The middle unfolding of \(A\), with indices \((a,b)|(c,d)\), has rank \(8\).
Both outer unfoldings also have rank \(8\). The finite identities needed below are
\begin{equation}\label{eq:gauge-short-identities}
\begin{aligned}
A_{aacd}&=0,& A_{abcc}&=0,\\
\sum_a\Omega_{aa}B_{daa}&=0,&
\sum_a\Omega_{aa}B_{aac}&=0,\\
\sum_a\Omega_{aa}A_{eaac}&=-9M_{ec},&
\operatorname{Tr}(\Omega N)&=-3780.
\end{aligned}
\end{equation}
These statements can be checked directly by substitution into
Eq.~\eqref{eq:gauge-real-coefficients}. Exact reconstruction of
\(G_0\) from the spin Hamiltonian verifies the ranks and
contractions. An independent Mathematica implementation constructs
the same tensors from the spin matrices.

For an explicit construction of the weight, replace the eight diagonal
entries of \(\Omega\) by unknowns \(w_a\). The two \(B\) contractions
and the \(A\) contraction in Eq.~\eqref{eq:gauge-short-identities}
form \(16+64\) linear equations, with the coefficient of \(M\)
fixed to \(-9\). Their coefficient matrix has rank eight. Exact
elimination gives the unique solution in
Eq.~\eqref{eq:gauge-pair-matrices}; contraction with \(N\) then gives
\(-3780\). An independent reconstruction of \(G_0,A,B,M,N\) from the spin
matrices verifies both this linear solve and the trace obstruction as a
polynomial identity in all eight components of \(v\). The tensors are rational
coefficients in the Hermitian basis above, without an additional
integer scaling. Thus the weight and the constant can be checked
without the polynomial-search calculation.

\subsection{Middle-cut bound for an arbitrary Hermitian current}

Group rows of the middle unfolding by whether the first index is zero, and
columns by whether the last index is zero. The resulting block matrix is
\[
\begin{pmatrix}X&Y_{\mathrm{full}}\\Z_{\mathrm{full}}&K\end{pmatrix}.
\]
The block \(K\) is \(A\) with zero rows and columns inserted for interior
identity indices, so \(\operatorname{rank}K=8\). Let \(N_R\) span its right
kernel and let the rows of \(N_L\) span its left kernel. Reduction of \(K\)
to rank normal form, followed by elimination of its nonsingular block, gives
\begin{equation}\label{eq:gauge-bordered-rank}
\rho_2\ge8+\operatorname{rank}(Y_{\mathrm{full}}N_R)
              +\operatorname{rank}(N_LZ_{\mathrm{full}}).
\end{equation}
Indeed, the remaining bordered matrix has a zero lower-right block, and its
rank is at least the sum of its two off-diagonal ranks.

The zero interior-index columns and rows of \(K\) select the submatrices
\(Y_{ab}=g_{0a0b}\) and \(Z_{ab}=g_{a0b0}\) of these residual blocks.
Since \(Y+Z=M\) is invertible,
\begin{equation}\label{eq:gauge-middle-bound}
\rho_2\ge8+\operatorname{rank}Y+\operatorname{rank}Z\ge16.
\end{equation}
The fixed outer unfoldings of \(A\) imply \(\rho_1,\rho_3\ge8\).
These bounds allow arbitrary Hermitian currents, not just scalar currents.

\subsection{The outer-rank-eight case}

Suppose \(\rho_1=8\). The eight active-first-index rows are independent because
their fixed-core restriction already has rank \(8\). There is therefore a
unique real vector \(v\) satisfying
\begin{equation}\label{eq:gauge-boundary-functional}
g_{0bcd}=\sum_{a=1}^8v_a g_{abcd},\qquad b,c,d=0,\ldots,8.
\end{equation}
The fixed-core zeros imply \(Y=0\), \(Z=M\), \(g_{00ab}=0\), and
\(g_{000a}=0\). The fixed orbit sums then give
\begin{equation}\label{eq:gauge-necessary-parameters}
\begin{aligned}
\ell_{abc}:=g_{abc0}&=B_{abc}-\sum_dv_dA_{dabc},\\
P_{ab}:=g_{ab00}&=N_{ab}-\sum_dv_d\ell_{dab},\\
u_a:=g_{a000},\qquad u&=-(P+M)^{\mathsf T}v,\qquad p=v^{\mathsf T}u=0.
\end{aligned}
\end{equation}
The formula for \(u\) uses the zero one-site orbit sums. The condition \(p=0\)
follows by applying Eq.~\eqref{eq:gauge-boundary-functional} to the all-zero
remaining indices, since \(g_{0000}=0\). Thus these equations are necessary
for every density with \(\rho_1=8\), rather than a selected family of gauges.

Delete the dependent first-index-zero rows of the middle unfolding and its
zero columns with third index zero and fourth index active. Order the remaining
rows as \((a,0),(a,b)\) and columns as \((c,0),(c,d),(0,0)\). The matrix is
\[
\begin{pmatrix}M&0&u\\\ell&A&\operatorname{vec}P\end{pmatrix},
\]
where \(\ell\) has shape \(64\times8\). Eliminating the invertible \(M\) gives
\begin{equation}\label{eq:gauge-rank-reduction}
\begin{aligned}
\rho_2&=8+\operatorname{rank}[A,\operatorname{vec}V],\\
V_{ab}&=P_{ab}-\sum_c\ell_{abc}(M^{-1}u)_c.
\end{aligned}
\end{equation}
The short identities in Eq.~\eqref{eq:gauge-short-identities} imply
\[
\sum_a\Omega_{aa}\ell_{daa}=0,\qquad
\sum_a\Omega_{aa}\ell_{aac}=9(Mv)_c,\qquad
\operatorname{Tr}(\Omega P)=-3780.
\]
Using the symmetry of \(M\), we obtain the obstruction
\begin{equation}\label{eq:gauge-weighted-obstruction}
\boxed{\operatorname{Tr}(\Omega V)+9p=-3780.}
\end{equation}
If \(\operatorname{vec}V\) were in the column space of \(A\), its diagonal
entries would vanish by \(A_{aacd}=0\), so \(\operatorname{Tr}(\Omega V)=0\).
Together with \(p=0\), this contradicts
Eq.~\eqref{eq:gauge-weighted-obstruction}. Hence \(\rho_2=17\).

The right outer-rank-eight case follows by reflection. Both outer ranks cannot
be \(8\), since they would force \(Y=Z=0\) despite \(Y+Z=M\ne0\).
If one is \(8\), the other is \(9\) and the middle rank is \(17\). If both are
\(9\), the middle rank is at least \(16\). Lemma~\ref{lem:graded-full-series}
therefore gives \(D\ge36\) in every case.

\subsection{Reflection-even bound and saturating constructions}

For a reflection-even Hermitian density, real coefficients satisfy
\(Z=Y^{\mathsf T}\). Since \(Y+Y^{\mathsf T}=M\) is negative definite,
\(Y\) is invertible: a real null vector \(x\ne0\) would give
\(x^{\mathsf T}Mx=0\), a contradiction. Equation~\eqref{eq:gauge-middle-bound}
then yields \(\rho_2\ge24\). Neither outer rank can be \(8\), because that
would force one of \(Y,Z\) to vanish. Thus \(D\ge2+9+24+9=44\).
Hermiticity is used in this real-matrix argument; no such reflection-even bound
is asserted for arbitrary complex non-Hermitian densities.

For the upper bounds, the normalized three-site partial trace of \(G_0\)
vanishes. Substitution of Eq.~\eqref{eq:gauge-partial-traces} into
Eq.~\eqref{eq:gauge-optimizing-densities} gives
\(\operatorname{Tr}_1G_{\mathrm L}=0\), while
\[
G_{\mathrm E}=G_0-T\otimes I^{\otimes2}
                    +2I\otimes T\otimes I-I^{\otimes2}\otimes T.
\]
This expression is reflection even. Both differences from \(G_0\) are explicit
three-site divergences and both densities are Hermitian SU(2) scalars.
Exact unfoldings give \((8,17,9)\) and \((9,24,9)\); TT insertion therefore
attains dimensions \(36\) and \(44\). The computational material also supplies
the rational MPO entries and the prefix/suffix lists for independent graded
Hankel witnesses of these dimensions. Their determinants equal the products of
the selected local unfolding minors, as in Lemma~\ref{lem:graded-full-series}.

\subsection{Individual tail minima}\label{app:gauge-tail}

For the two optimizing realizations, let \(R_4,C_4\) denote matrices spanning
the exact-length-four reachable and observable spaces. Exact calculation gives
\begin{center}
\begin{tabular}{@{}lrrrr@{}}
\toprule
Density & \(\dim R_4\) & \(\dim C_4\) & \(\operatorname{rank}(R_4C_4)\) & Tail bound\\
\midrule
\(q_{\mathrm L}\) &36&26&26&36\\
\(q_{\mathrm E}\) &35&35&26&44\\
\bottomrule
\end{tabular}
\end{center}
Both spaces equal their length-five counterparts, verified by the corresponding
union ranks. Transition-space induction then gives stability at every longer
length. The restriction-dimension argument of Theorem~\ref{thm:tail-rigidity}
gives a lower bound
\(\dim R_4+\dim C_4-\operatorname{rank}(R_4C_4)\), namely \(36\) or \(44\).
The full-series realizations attain these bounds. This proves the two individual
tail minima in Table~\ref{tab:gauge-optima}; it does not optimize the tail-only
problem over \(\mathcal G_4\) or \(\mathcal G_4^+\).

The obstruction in Eq.~\eqref{eq:gauge-weighted-obstruction} is also verified
as an exact polynomial identity in eight formal variables. This symbolic
check supplements the displayed trace proof; it is not a numerical scan.
The finite tensors and tail-space pairings are supplied in the Supplementary
Computational Material.

\section*{Data and code availability}
Reference~\cite{ZhouYu2026SCM} identifies the computational release.
The exact-arithmetic code, machine-readable data, and reproduction
instructions are provided in the Supplementary Computational Material.
Repository record for the release (DOI):
\url{https://doi.org/10.5281/zenodo.24102750}.

\section*{Declaration of generative AI and AI-assisted technologies in the
manuscript preparation process}
During the preparation of this work, the authors used large language
models to assist with language editing of the manuscript and review of the
computational code. After using this tool, the authors reviewed and edited
the content as needed and take full responsibility for the content of the
published article.

\bibliographystyle{elsarticle-num-aop}
\bibliography{references}

\end{document}